\documentclass[sigconf]{acmart}

\setcopyright{none}              

\usepackage{tabularx, colortbl}
\usepackage{multirow, makecell}
\usepackage{enumitem}
\usepackage[T1]{fontenc}
\usepackage{stfloats}
\usepackage{array}
\usepackage{listings}
\usepackage{amsmath}
\usepackage{xspace}

\newcommand{\eg}{\textit{e.g.}}

\newcommand{\CoolName}[1]{Generative Proxy}
\newcommand{\quot}[1]{\emph{``#1''}}
\newcommand{\final}[1]{{\color{black}{#1}}}

\makeatletter
\def\@subsubsecfont{\sffamily\bfseries}
\makeatother

\copyrightyear{2026}
\acmYear{2026}
\setcopyright{cc}
\setcctype{by-nc-nd}
\acmConference[UIST '26]{The 39th Annual ACM Symposium on User Interface Software and Technology}{November 02--05, 2026}{Detroit, MI, USA}
\acmBooktitle{The 39th Annual ACM Symposium on User Interface Software and Technology (UIST '26), November 02--05, 2026, Detroit, MI, USA}
\acmDOI{10.1145/3830398.3830482}
\acmISBN{979-8-4007-2856-3/2026/11}

\begin{document}

\title{Generative Proxy: Synthesizing Proxy-Based Interfaces for Real-World Interaction Across AR Glasses}



\author{Xianhao Carton Liu}
\email{liu03008@umn.edu}
\orcid{0009-0006-3528-9651}
\affiliation{%
  \institution{University of Minnesota}
  \city{Minneapolis}
  \state{Minnesota}
  \country{USA}}

\author{William Chastek}
\email{chast068@umn.edu}
\orcid{0009-0005-8136-8935}
\affiliation{%
  \institution{University of Minnesota}
  \city{Minneapolis}
  \state{Minnesota}
  \country{USA}
}

\author{Eric J Gonzalez}
\email{ejgonz@google.com}
\orcid{0000-0002-2846-7687}
\affiliation{%
  \institution{Google}
  \city{Seattle}
  \state{Washington}
  \country{USA}
}

\author{Mar Gonzalez-Franco}
\email{margon@google.com}
\orcid{0000-0001-6165-4495}
\affiliation{%
  \institution{Google}
  \city{Seattle}
  \state{Washington}
  \country{USA}
}

\author{Chen Zhu-Tian}
\email{ztchen@umn.edu}
\orcid{0000-0002-2313-0612}
\affiliation{%
  \institution{University of Minnesota}
  \city{Minneapolis}
  \state{Minnesota}
  \country{USA}
}

\renewcommand{\shortauthors}{Liu et al.}

\begin{abstract}
Interacting with real-world objects in AR is difficult, \final{especially when targets are distant, cluttered, or occluded}. These challenges are amplified on emerging lightweight AR glasses, which often lack binocular or large field of view on display, but also continuous inputs, such as hand or eye tracking. Proxy-based interfaces offer an alternative by allowing users to interact with virtual abstractions of physical objects that can be repositioned, reorganized, and adapted to the task and device. However, designing such interfaces is currently manual and highly device-specific. We present Generative Proxy, a method for automatically generating proxy-based interfaces from three specifications: scene, intent, and device capabilities. We formulate generation as a constrained synthesis problem that first produces valid interfaces for the target device and task, then ranks candidates using semantic and articulatory distance inspired by direct manipulation theory. We demonstrate Generative Proxy across diverse scenes, device profiles, and user intents.
Expert evaluation shows initial evidence that generated proxy UIs are useful and usable, highlighting proxy-based abstraction as a promising interaction paradigm for future AR glasses.
\end{abstract}

\begin{teaserfigure}
    \centering
  \includegraphics[width=0.98\textwidth]{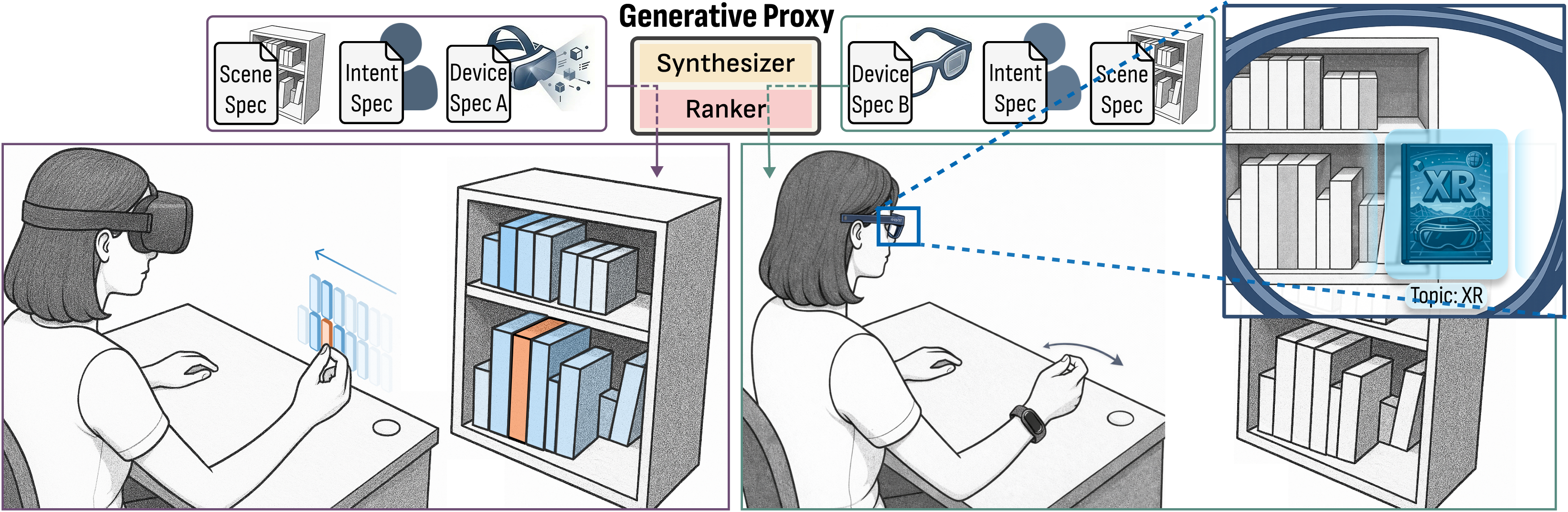}
  \caption{Generated proxy UIs for immersive AR headsets (left, \final{adapted from~\cite{liu2025reality}}) and lightweight AR glasses (right). Generative Proxy accounts for diverse scenes, user intents, and device profiles to produce appropriate proxy UIs. As shown on right, proxy UIs enable interaction with real-world objects on lightweight AR glasses, despite the lack of ray casting or pointing input.}
  \Description{.}
  \label{fig:teaser}
\end{teaserfigure}


\maketitle
\section{Introduction}

Recent advances in glasses-like wearable computing suggest a shift from bulky immersive headsets toward lighter, more wearable augmented reality (AR) form factors. 
Companies including Google, Meta, and Snap are actively developing glasses-based platforms, reflecting growing interest in AR experiences embedded in everyday interaction with the physical world~\cite{suzuki2025everyday}.
As these devices become more practical and socially acceptable, a central challenge is how best to facilitate interaction with the real-world around the user.

A common approach in AR is to interact with physical objects directly using ray-casting, gaze, or other remote targeting techniques~\cite{pfeuffer2024design, gonzalez2024guidelines}.
However, direct interaction is constrained by the physical scene itself: \final{objects may be distant, cluttered, occluded, or awkward to access~\cite{whitlock2018interacting}}.
Moreover, it is often a poor fit for lightweight AR glasses, whose input capabilities may be limited to voice or discrete gestures rather than continuous mid-air pointing~\cite{lee2020ubipoint}.


\final{\emph{Proxy-based interfaces}}~\cite{liu2025reality} offer an \final{complementary paradigm to direct interactions}. 
Instead of selecting or manipulating physical objects directly, 
users interact with virtual surrogates that represent them.
Because these surrogates are decoupled from the objects' physical arrangement, they can be repositioned, resized, and reorganized to better support the user's task. 
For example, rather than requiring a user to target multiple books, a proxy can group relevant books into a single representation that is easier to inspect and select.
\final{Proxy-based interfaces are therefore especially useful when targets are distant, cluttered, occluded, visually similar, or otherwise difficult to reference unambiguously.}

However, proxy-based interfaces remain largely handcrafted and device-specific. 
AR devices vary widely in display and input capabilities, from spatially registered 3D content to 2D overlays, and from full hand tracking to gaze or discrete gestures~\cite{liu2026can, milgram1994taxonomy, laviola20173d, nguyen2023hand}. 
As a result, a proxy UI designed for spatial 3D presentation with continuous hand input (\autoref{fig:teaser} Left) and may not be appropriate for another device with only screen-space output and discrete gestures (\autoref{fig:teaser} Right). 
Manually redesigning and retuning such interfaces for every device therefore does not scale.

In this work, we ask: \emph{how can proxy-based interfaces be generated automatically from the physical scene, the user's intent, and the target device capabilities?} 
We present \CoolName{}, a method for synthesizing proxy-based interfaces for real-world interaction across heterogeneous AR capability conditions. 
Given typed specifications of scene, intent, and device capability, \CoolName{} first synthesizes interface candidates that are valid for the target task and platform, then ranks them using direct-manipulation-inspired criteria that favor interfaces with lower semantic, articulatory, and evaluation distance~\cite{hutchins1986direct}. 
This yields proxy interfaces that are both device-compatible and better aligned with the user's task.

We demonstrate \CoolName{} through a proof-of-concept implementation and three application scenarios, and further probe the pipeline through an internal ranking sanity check and an expert evaluation with experienced XR users ($N=12$). 
Together, these studies provide initial evidence that proxy-based UIs can support interaction with real-world objects across a broader range of AR capability conditions, and that our method can generate plausible and usable proxy UIs across varying device capabilities, scenes, and intents.
In summary, this paper makes the following contributions:

\begin{itemize}
    \item \textbf{Design.} A formal specification framework for representing proxy interfaces, physical scenes, user intents, and device capabilities, together with a constrained synthesis-and-ranking pipeline for generating proxy interfaces in AR.
    \item \textbf{Prototype.} A proof-of-concept cross-device implementation and three application scenarios demonstrating how proxy-based UIs can be generated and applied across diverse AR capability profiles.
    \item \textbf{Evaluation.} Initial evidence from an internal ranking sanity check and an expert evaluation (N=12) probing the usefulness of generated proxy UIs, how device-specific variants are perceived, and where the current pipeline breaks down.
\end{itemize}

\section{Related Work}

This work concerns interacting with real-world objects in AR for digital tasks. 
We review prior work in three areas: interaction with real-world objects in AR, UI synthesis and generation, and contextually adaptive AR interfaces.

\subsection{Interacting with Real-World Objects in AR}
We use \emph{interacting with real-world objects in AR} to describe systems that treat physical objects as referents for associated digital information or actions. 
We organize this literature through a hardware- and interaction-driven lens, since AR platforms have strongly shaped how such interactions are realized.

\paragraph{\final{Handheld and screen-mediated interactions.}}
Early AR systems typically accessed physical objects through the device itself, for example via fiducial markers~\cite{kato1999marker} or handheld mobile screens~\cite{henrysson2005face}. 
Marker-based systems~\cite{kan2009applying, ahuja2019lightanchors} attached digital information and actions to instrumented referents through explicit tags, while mobile AR systems such as \emph{RealitySketch}~\cite{suzuki2020realitysketch} supported object-linked content through touchscreen interaction. 
In these systems, the device remained the primary locus of interaction.

\paragraph{Head-worn embodied interactions.}
With head-worn AR, interaction shifted from the device toward embodied spatial input. 
Systems in this space use gaze, head motion, hand rays, gestures, voice, and direct touch-like interaction to specify physical targets and invoke digital actions. 
A recurring theme is multimodal disambiguation: \emph{Pinpointing}~\cite{kyto2018pinpointing} combines head motion and gaze for precise AR selection, \emph{GazePointAR}~\cite{lee2024gazepointar} combines gaze, gesture, and speech for situated reference resolution, and \emph{Uncertain Pointer}~\cite{tsai2026uncertain} visualizes referential ambiguity during AR target selection. 
A parallel line of work treats everyday objects as interfaces~\cite{chen2020augmenting,tong2022exploring,du2022opportunistic,monteiro2023teachable,hettiarachchi2016annexing}. 
For example, \emph{XR-Objects}~\cite{dogan2024augmented} uses multimodal large language models to let real objects expose context-sensitive menus and actions without preregistration. 
Together, these systems move beyond device mediation toward embodied, object-grounded interaction.

\paragraph{Lightweight wearable and distributed interactions.}
Lightweight AR glasses introduce a different regime of constraints, including limited sensing fidelity, restricted feedback bandwidth, and reduced support for precise manual input~\cite{lee2023embodied}. \final{To address these limitations, Ren et al. developed a peripheral AR display technique to facilitate real-world object selection on lightweight AR glasses ~\cite{ren2026periphar}.} 
More broadly, recent work increasingly relies on distributed sensing and AI-mediated inference to recover user intent from partial signals.
These systems combine cues such as speech~\cite{miniotas2006speech}, coarse pointing~\cite{lee2024gazepointar}, gaze~\cite{chatterjee2015gaze}, wrist- or ring-based input~\cite{chatterjee2025flowring}, and egocentric scene context~\cite{dogan2024augmented} with learned models to infer the intended object or action.
Examples include SensibleAgent~\cite{lee2025sensible} for context-aware interaction and Squiggle~\cite{fashimpaur2025squiggle} for inferring multi-object reference through implicit lasso input. 
We do not review the broader literature on wearable gesture and activity recognition~\cite{caramiaux2015understanding, wang2025computing}, which is adjacent but not central here.

\paragraph{Summary.}
Across these lines of work, interaction typically remains tied to directly referring to, touching, or pointing at the target object itself.
Proxy-based interfaces redirect interaction from the physical referent to an abstract or synthesized proxy that preserves its digital semantics while reducing the need for direct access to the object. 
\final{\emph{RealityProxy}~\cite{liu2025reality} is a prior instantiation of this broader interaction paradigm.
Building on this foundation, our work asks how proxy interfaces can be synthesized across heterogeneous AR systems with varying sensing, ergonomic, and form-factor constraints.
}

\vspace{-2mm}
\subsection{User Interface Synthesis and Generation}
The common premise of prior work on UI generation is that interfaces can be derived from representations more abstract than the final UI, such as task models, declarative specifications, or natural-language descriptions. They differ mainly in how much structure is provided in advance and how much control remains over the generation process.

\paragraph{Model-based user interfaces.}
Model-based UI (MBUI) approaches derive interfaces from abstract models such as task models, domain models, and mappings from abstract interactions to concrete widgets~\cite{szekely1993beyond,zanden1990automatic,nichols2004improving}. 
This separation supports systematic generation, retargeting, and consistency. 
More recent examples include ORC Layout~\cite{jiang2019orc}, which formulates adaptive GUI layout with OR-constraints, and Scout~\cite{swearngin2020scout}, which supports mixed-initiative exploration of layout alternatives through high-level constraints. 
Yet, MBUI often requires substantial upfront modeling and tends to assume relatively static, developer-authored structures.

\paragraph{Specification-based generation.}
Specification-based generation can be seen as a more constrained form of model-based generation, centered on structured intermediate representations that can be interpreted reliably and sometimes manipulated directly. 
For example, Bespoke~\cite{vaithilingam2019bespoke} synthesizes GUIs for command-line workflows from demonstrations, while Vega-Lite~\cite{satyanarayan2016vega} provides a declarative grammar for visualization specification and systems such as DynaVis~\cite{vaithilingam2024dynavis} build persistent editing interfaces on top of such representations. 
The main advantage is interpretability and control: explicit primitives and mappings make the generation space more reusable and consistent. 
The tradeoff is that expressiveness is bounded by the underlying specification language.

\paragraph{AI-based constraint-guided generation.}
Recent work has pushed UI generation toward direct synthesis from natural-language prompts and other high-level inputs using large language models. This lowers authoring barriers and broadens the range of tasks that can be addressed, but direct prompt-to-code generation remains difficult to control and revise~\cite{chen2021evaluating}. Systems such as UICoder~\cite{wu2024uicoder}, LLM-assisted design support in Figma~\cite{figma_design_2026}, Misty~\cite{lu2025misty}, and Squire~\cite{leung2025squire} respond by reintroducing structure around AI generation through feedback, intermediate representations, and mixed-initiative workflows. Related work on malleable interfaces~\cite{cao2025generative, min2025malleable} further treats AI as a mechanism for continued interface evolution rather than only initial generation. Together, this literature suggests that AI expands the scope of UI generation, but benefits from structured intermediates, explicit constraints, and mixed-initiative interaction.

Building on these threads, we adopt a hybrid specification-and-AI approach: specification-based synthesis constrains the design space to valid candidates, while a theory-grounded LLM-based ranker prioritizes \final{interfaces with lower semantic, articulatory, and evaluation distance}.

\subsection{Contextually Adaptive Interfaces in AR}

Adaptive interfaces have long been central to AR because decisions about what to show, where to place it, and how it should appear depend on the user's immediate physical context~\cite{grubert2016towards}. 
A large body of work has therefore focused on adaptation grounded in environmental geometry~\cite{nuernberger2016snaptoreality, DBLP:journals/tvcg/ChenS0WQW20, DBLP:conf/uist/FenderLHA017, DBLP:conf/chi/FenderHA018}. 
For example, AdapTUI~\cite{he2024adaptui} formulates adaptation as an optimization problem that jointly considers geometric and human factors for tangible AR interfaces.

More recent work extends adaptation beyond geometry by incorporating semantics of the environment. 
Lindlbauer et al.~\cite{lindlbauer2019context} adapt placement and level of detail based on contexts and user cognitive load; Tahara et al.~\cite{DBLP:conf/ismar/TaharaSNI20} use scene graphs to maintain consistency across locations; AdapTutAR~\cite{DBLP:conf/chi/HuangQWPSCRQ21} personalizes AR instructions to user characteristics; SemanticAdapt~\cite{cheng2021semanticadapt} uses computer vision to attach semantically relevant augmentations; and Zhu-Tian et al.~\cite{zhu2023rl} apply reinforcement learning to continuously adapt label layouts. 
Authoring tools such as CAPturAR~\cite{wang2020capturar}, AUIT~\cite{evangelista2022auit}, and ScalAR~\cite{qian2022scalar} further lower the barrier to building adaptive AR applications.

Our work shares this interest but differs in where adaptation enters the pipeline. 
Rather than adapting pre-authored interfaces online, we cast the problem as one of UI synthesis from device capability, user intent, and scene structure. 
The resulting interfaces could subsequently be combined with runtime adaptation techniques for further refinement of layout and placement.

\section{Problem Formulation}

Proxy interfaces has the potential to facilitate real-world interactions across heterogeneous AR capability conditions, each with different sensing, input, and output capabilities. 
However, manually redesigning a proxy interface for every device is therefore impractical and does not scale. 
This motivates the need for a method that can generate proxy interfaces automatically while remaining appropriate to the user, the environment, and the target device.
From this problem setting, we derive two requirements:

\paragraph{\textbf{R1: Generating Valid Interfaces from Scene, Intent and Device Capability.}}
Whether a proxy interface is valid depends on three inputs: 
(1) \emph{scene}, which specifies the available physical objects and their relevant attributes; 
(2) \emph{intent}, which captures what the user is trying to do; and 
(3) \emph{device capability}, which defines the target platform's available input and output. 
A generated proxy interface is valid only if it is compatible with the device capabilities and supports the intended interaction over the given scene.

\paragraph{\textbf{R2: Ranking Valid Interfaces by Directness of Interaction.}}
Validity alone is insufficient. 
For a given scene, intent and device, many proxy interfaces may satisfy these basic constraints, yet differ substantially in usability. 
The remaining problem is therefore not only to generate feasible interfaces, but also to determine which ones better preserve the feel of direct manipulation.

We frame this distinction using Hutchins, Hollan, and Norman's account of direct manipulation~\cite{hutchins1986direct}, which relates perceived directness to the gulfs of execution and evaluation. 
In particular, the gulf of execution depends in part on \emph{semantic distance}, or how closely the available actions correspond to the user's intended task, and \emph{articulatory distance}, or how easily those actions can be physically expressed.
This perspective is especially relevant for proxy-based interfaces because they decouple interaction from the original physical arrangement of objects. 
That decoupling creates opportunities to reorganize interactive representations in ways that better match the user's task and the device's interaction constraints. 
For example, if a user wants to select all HCI books, a proxy could group them into a single nearby representation, reducing semantic distance by aligning the interface with the task structure and reducing articulatory distance by enabling a simple within-reach action.

Accordingly, beyond generating valid proxy interfaces, the problem also requires prioritizing those that reduce semantic and articulatory distance for the intended interaction.


\section{\CoolName{}: Approach}

\begin{figure*}[b]
    \centering
    \includegraphics[width=1\textwidth]{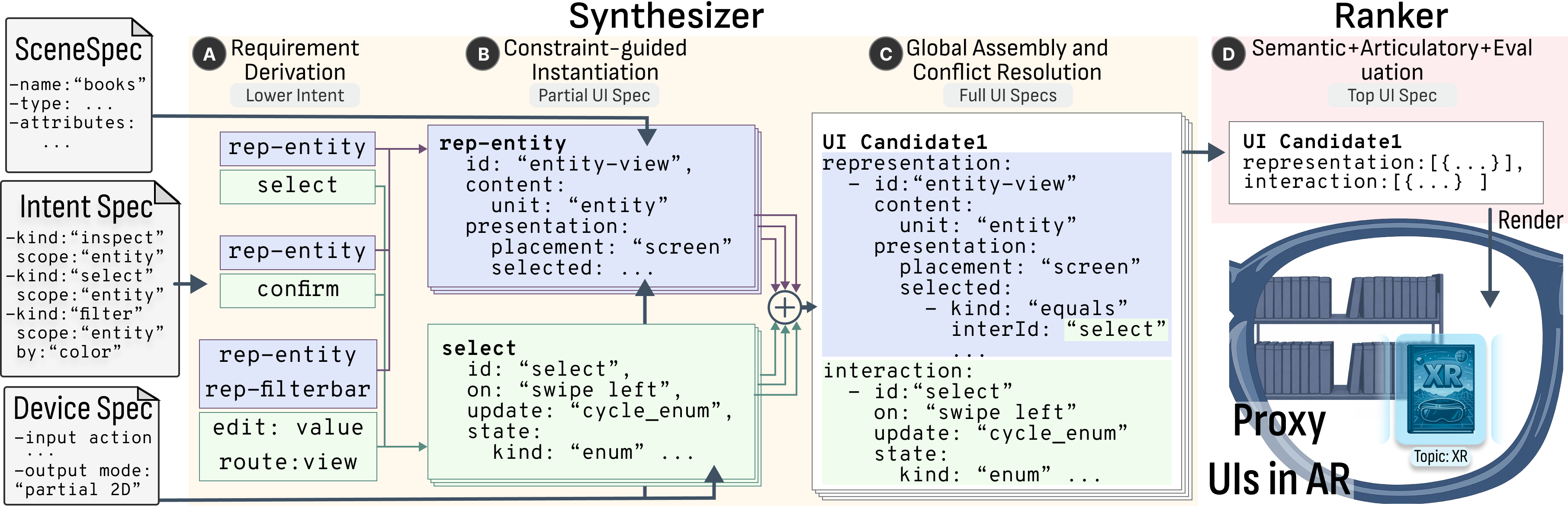}
    \vspace{-6mm}
    \caption{\final{Pipeline overview with a running bookshelf example.} From scene, intent, and device specifications, the synthesizer derives requirements, instantiates locally valid representation (``rep'') and interaction components under scene and capability constraints, assembles them into full UI candidates, and the ranker selects the most direct candidate using semantic, articulatory, and evaluation distance.
    }
    \label{fig:pipeline}
\end{figure*}

We formulate proxy interface generation as a constrained synthesis problem.
Given a scene specification, a user intent specification, and a device capability specification,
the system synthesizes candidate interfaces and ranks feasible candidates using direct-manipulation-inspired distances. 
\final{~\autoref{fig:pipeline} illustrates the pipeline through a running bookshelf example for partial display AR glasses.}

\subsection{Typed Specifications}
We define a typed specification for the scene, user intent, device capabilities, and proxy interfaces.
Together, these specifications make the synthesis space explicit and provide contracts for what is present in the environment, what the user wants to do, what the device can support, and what counts as a valid interface.
Below we summarize each specification and its design rationale; \final{full definitions are provided in the supplementary materials.}

\subsubsection{Scene}

We represent the \texttt{Scene} as a set of typed entities:
\[
\mathrm{Scene} := \{(\texttt{name}, \texttt{type}, \texttt{attributes}, \texttt{pos}, \texttt{parent?})\}
\]

Each entity records its identity, semantic type, attributes, approximate position, and optional parent relation, similar to the abstractions used in Reality Proxy~\cite{liu2025reality}. 
In practice, this level of abstraction is sufficient for deciding which objects can be proxied together, which objects share meaningful structure, and how a generated interface should remain grounded in the real world.
\final{For example, the input \texttt{SceneSpec} in \autoref{fig:pipeline} describes bookshelf entities with attributes such as topic and color.}

\subsubsection{User Intents}
\vspace{-2mm}
The user's goal is described by an \texttt{IntentSpec}:
\[
\mathrm{IntentSpec} := (\texttt{kind}, \texttt{scope}, \texttt{by?})
\]

\texttt{kind} specifies the operation to be performed, \texttt{scope} specifies what it ranges over, and \texttt{by} optionally refines that operation, for example by grouping or filtering criterion.
A key design decision is that \texttt{IntentSpec} captures \emph{what} the user wants to accomplish without prescribing \emph{how} the interface should realize it.
This separation is essential for synthesis: the same high-level goal, such as inspecting distant objects or reorganizing a cluttered set, may be realized by very different proxy interfaces depending on the scene and the target device.
\final{In this work, we implemented five intents \final{(i.e., \emph{inspect, select, organize, filter, and relevel})} based on interaction taxonomies from prior work~\cite{laviola20173d} and the operations supported in Reality Proxy~\cite{liu2025reality}, although other choices are possible.}
\final{The example \texttt{IntentSpec} in \autoref{fig:pipeline} requests inspect, select, and filter.}

\subsubsection{Device Capabilities}

A \texttt{CapabilitySpec} captures the interaction resources available on a target device:
\[
\begin{aligned}
    \mathrm{CapabilitySpec} &:= (\{\mathrm{InputAction}\}, \mathrm{OutputMode})\\
    \mathrm{InputAction} &:= (\texttt{name}, \texttt{kind}, \texttt{dims}, \texttt{frame})\\
    \mathrm{OutputMode} &\in \{\emph{world}, \emph{screen}, \emph{partial screen}, \emph{non-visual}\}
\end{aligned}
\]

\texttt{InputAction} abstracts an available input by its type, dimensionality \final{(i.e., \emph{3D}, \emph{2D}, and \emph{Symbolic})}, and reference frame \final{(i.e., \emph{world}, \emph{screen}, and \emph{none})}, while \texttt{OutputMode} describes how content can be presented.
The choice of fields and values is informed by prior taxonomies of AR interfaces~\cite{foley1996computer} and interactions~\cite{laviola20173d}. 
The design intentionally omits hardware-specific details and retains only the properties that constrain synthesis: 
how expressive input is, how it is spatially grounded, and where output can appear.
This allows the synthesizer to reason over heterogeneous devices at the level relevant to interface generation rather than device hardware.

\subsubsection{Interface}
An \texttt{InterfaceSpec} is where the three input specifications come together. It determines how a scene is transformed into proxy-based content and presented on a given device, and how the user can act on it through the device's input resources:
\[
\begin{aligned}
\mathrm{InterfaceSpec} &:= (\{\mathrm{Representation}\}, \{\mathrm{Interaction}\})\\
\mathrm{Representation} &:= (\texttt{id}, \texttt{content}, \texttt{presentation}, \texttt{active?})\\
\mathrm{Interaction} &:= (\texttt{id}, \texttt{on}, \texttt{update}, \texttt{state}, \texttt{active?})
\end{aligned}
\]
Inspired by Vega-Lite~\cite{satyanarayan2016vega}, we adopt a declarative design for the user interfaces: 
A \texttt{Representation} separates \texttt{content}, which specifies \emph{what to show}, from \texttt{presentation}, which specifies \emph{how to show it}; an \texttt{Interaction} specifies how device inputs update interface state over time.
\final{The \texttt{content} may expose one or more entities or groups formed by an attribute. 
The \texttt{presentation} may arrange these proxies according to their content structure or interaction requirements and place them in world space, screen space, or omit visual placement. 
Together, these choices support alternatives such as world-anchored entity proxies, screen-space grouped views, and symbolic lists operated through discrete input.}

We separate representation from interaction so the synthesizer can reason independently about what is shown and how it can be manipulated, while making compatibility constraints explicit.
This declarative structure supports checking whether a proposed interface is realizable on the target device.


\subsection{Intent-Driven Synthesis}
Rather than mapping inputs to a single predetermined output, the synthesizer solves a constrained synthesis problem: the same scene, intent, and device may admit multiple valid proxy interfaces. 
Our key observation is that the three specifications play different roles in this process. 
\texttt{Scene} determines what real-world structure is available to expose or transform, \texttt{IntentSpec} determines what semantic structure and operations the interface must support, and \texttt{CapabilitySpec} determines which realizations are feasible on the target device. 
This decomposition lets us formulate synthesis as intent-driven compilation under scene and device constraints.

\vspace{-2mm}
\paragraph{Requirement derivation from intent.}
The synthesizer begins by translating each \texttt{IntentSpec} into a small set of canonical requirements,
\final{such as deriving an entity-level representation together with operations like selection or confirmation (\autoref{fig:pipeline}a)}.
This step separates user goals from interface realization. 


\begin{figure}[h]
    \centering
    \vspace{-2mm}
    \includegraphics[width=0.9\linewidth]{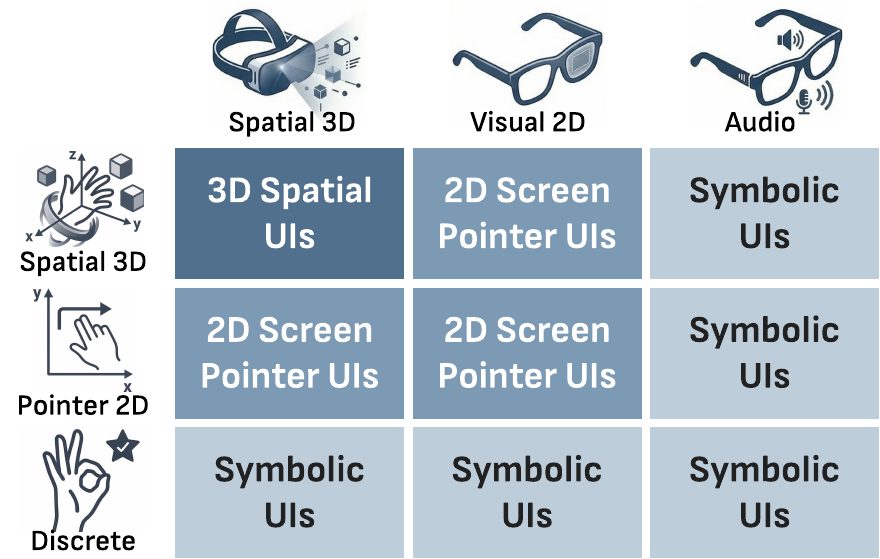}
    \vspace{-2mm}
    \caption{Interface realizability depends jointly on input and output capability. Different combinations of input and output modality admit different interface families.}
    \label{fig:matrix}
    \vspace{-4mm}
\end{figure}

\paragraph{Capability-constrained instantiation.}
The synthesizer then instantiates candidate representations and interactions that satisfy these requirements while remaining compatible with both the scene and the device. 
Scene structure constrains what content transformations are admissible: grouping requires meaningful attributes, hierarchy operations require hierarchical structure, and proxy content must correspond to entities that actually exist. 
Device capability constrains how that content can be realized and manipulated. 
As illustrated in Figure~\ref{fig:matrix}, interface form depends jointly on input and output capability: richer spatial input/output combinations support 3D spatial UIs, whereas more limited combinations favor screen pointer-based or symbolic UIs. 
\final{In \autoref{fig:pipeline}b, this stage produces locally valid partial UI specifications.}

\vspace{-1mm}
\paragraph{Global assembly and graceful degradation.}
Locally valid components may still conflict when combined into a complete interface. 
Multiple intents may compete for the same input actions, induce redundant representations, or require coordinated alternative views. 
We therefore formulate assembly as a bounded search over locally valid candidates, selecting combinations that maximize intent coverage while respecting device constraints and avoiding conflicts. 
\final{As shown in \autoref{fig:pipeline}c, the result of this stage is a set of full UI candidates rather than a single interface.}
This final stage also enables graceful degradation: when all requested intents cannot be realized simultaneously, the synthesizer returns interfaces that satisfy the largest feasible subset rather than failing outright.

The output is a set of candidate \texttt{InterfaceSpec}s that satisfy the user's intents for the current scene and device. Further implementation details are provided in an open-source repository \url{https://github.com/sys3-lab/GenerativeProxy/}.

\subsection{Ranking Candidate Interfaces}

Synthesis may produce multiple valid \texttt{InterfaceSpec}s for the same scene, device, and intents. 
We therefore rank candidates (\final{\autoref{fig:pipeline}d}) by operationalizing direct manipulation as three distances: \emph{semantic execution distance}, \emph{articulatory execution distance}, and \emph{evaluation distance}. 
Intuitively, preferred interfaces expose the right task structure, make the required actions easy to perform, and make their effects easy to interpret. 
For each candidate, we aggregate these distances across intents using priority weights, so interfaces that better support higher-priority intents are ranked more highly.

\vspace{-1mm}
\paragraph{Semantic execution distance.}
Semantic execution distance measures whether an interface exposes the representation needed for an intent at the appropriate semantic level. 
We compute it by comparing the intent with the synthesized \texttt{Representation.content}. 
Interfaces that make the required structure explicit receive lower cost; those that require the user to work through a less suitable representation receive higher cost. 
For example, an intent to organize objects by category is better supported by a category-grouped proxy view than by a flat list of entities.

\vspace{-1mm}
\paragraph{Articulatory execution distance.}
Articulatory execution distance measures how difficult it is to carry out an intent with a candidate interface. 
We compute this distance using two components: a rule-based interaction cost and an LLM-based ergonomic score. 
The rule-based component captures factors directly observable from the specification, including the number of steps, required view or mode switches, input precision, and penalties for missing operations. 
The LLM component captures residual ergonomic properties that are difficult to formalize symbolically, such as whether an action mapping feels natural on the intent and whether the overall control scheme is coherent and discoverable. 
This decomposition keeps the ranking interpretable while allowing articulatory quality to reflect both measurable burden and higher-level ergonomic plausibility.

\vspace{-1mm}
\paragraph{Evaluation distance.}
Evaluation distance measures how directly an interface makes action outcomes legible. We favor feedback embedded in, or tightly coupled to, the affected scene entities; for example, highlighting a selected object is better than showing a separate abstract state indicator. We compute this term using a rule-based tiered scheme derived from the device capability matrix (\autoref{fig:matrix}): in-scene feedback receives the lowest cost, while screen pointer-based or symbolic feedback receives progressively higher cost as it becomes more displaced from the relevant scene object. Thus, evaluation distance reflects the spatial and semantic gap between feedback and its referent.

Finally, we apply a small shortlist-only LLM coherence penalty to top-ranked candidates to capture whole-interface inconsistencies that are difficult to detect from individual interactions in isolation. 
Overall, the ranker combines deterministic task- and capability-based scoring with targeted LLM judgment for ergonomic plausibility, allowing it to prefer interfaces that are not only valid, but also more direct in practice.

\section{Implementation}

We implement \CoolName{} as a proof-of-concept technique probe with three components: 
a synthesis-and-ranking backend, 
a runtime pipeline for acquiring scene specifications, and 
device-specific runtimes that render synthesized interfaces and bind them to available input/output capabilities.
\final{
In the current implementation, \texttt{IntentSpec} and \texttt{CapabilitySpec} are specified in advance by developers, whereas \texttt{SceneSpec} is acquired at runtime when the user invokes proxy generation. This separation gives developers control over the supported intent vocabulary and target devices while allowing the generated interface to adapt to the current scene.}

\paragraph{Synthesis and ranking backend.}
We implement the synthesizer and ranker in TypeScript as a Node.js web application. 
Both operate over JSON-based specifications: \texttt{IntentSpec} and \texttt{CapabilitySpec} are provided in advance, while \texttt{Scene} is acquired at runtime. 
The backend combines deterministic rules with LLM calls. 
Rule-based logic handles specification validation, candidate synthesis, compatibility checking, and parts of distance scoring, while LLM calls are reserved for judgments that are difficult to formalize symbolically, such as ergonomic plausibility and semantic enrichment of scene entities. 
The backend outputs synthesized \texttt{InterfaceSpec}s as JSON for device-specific runtimes.

\begin{figure}[h]
    \centering
    \includegraphics[width=1\linewidth]{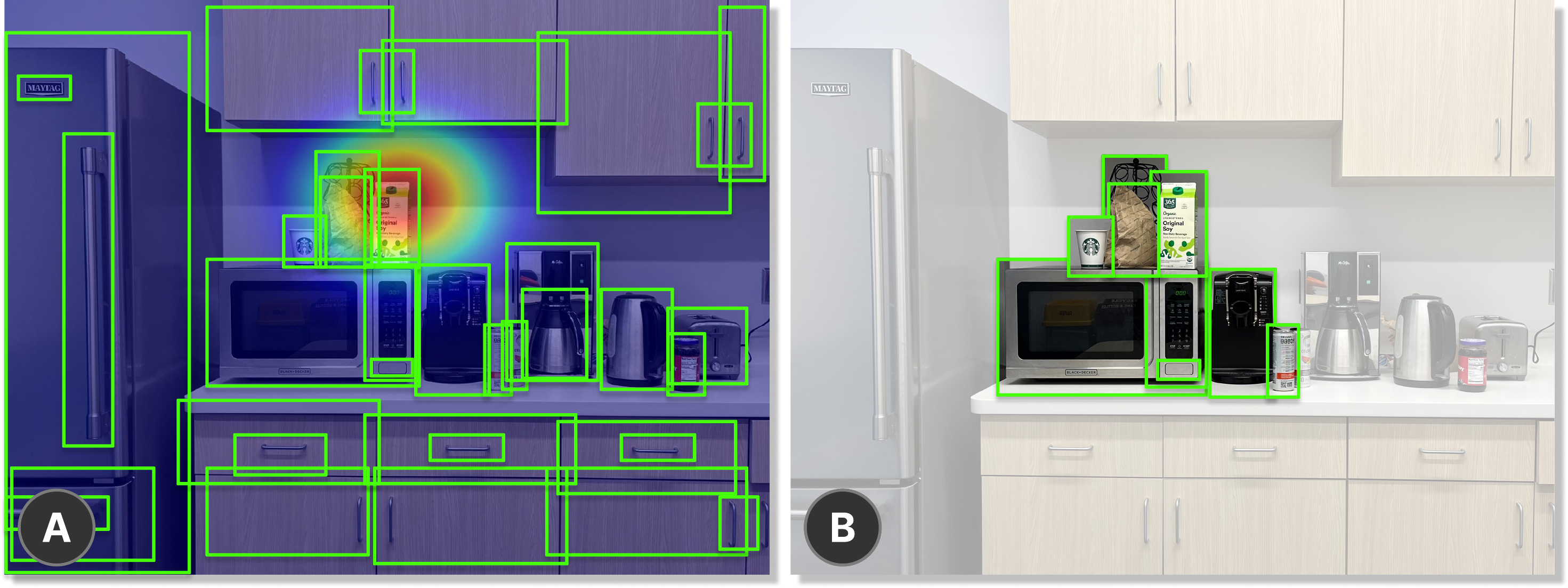}
    \vspace{-6mm}
    \caption{Gaze estimation narrows the scene for synthesis by filtering detected objects. (A) Gaze heatmap. (B) Objects retained after filtering.}
    \label{fig:gazebbox}
    \vspace{-4mm}
\end{figure}

\paragraph{Runtime scene specification acquisition.}
Because proxy synthesis depends on the current environment, the scene specification must be constructed online. 
Following Reality Proxy~\cite{liu2025reality}, we begin with object detection over the live egocentric view when the user triggers proxy UI generation. 
Whereas the original pipeline filters detections using eye gaze, many lightweight AR devices do not provide direct eye tracking or reliable head-gaze input. 
We therefore replace direct gaze sensing with a state-of-the-art gaze prediction model~\cite{lai2024eye} that estimates gaze region from a short sequence of head-motion images. 
We intersect the predicted gaze region with detected bounding boxes to obtain the subset of objects used to construct the \texttt{Scene} specification (\autoref{fig:gazebbox}), retaining only boxes intersecting regions whose predicted gaze score exceeds 0.75. 
We further use LLMs (GPT 5.5) to enrich detections with semantic attributes and recursive object detection to recover containment hierarchy.

\begin{figure}[h]
    \centering
    \includegraphics[width=1\linewidth]{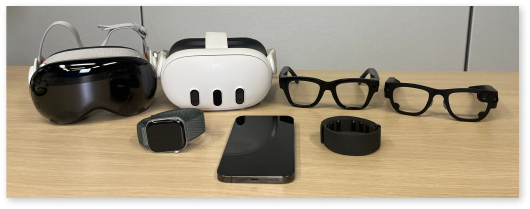}
    \vspace{-6mm}
    \caption{Different AR display and input devices used in our prototypes. From left to right, the first row shows Apple Vision Pro, Meta Quest 3, Meta Ray-Ban Display, and Meta Project Aria glasses; the second row shows Apple Watch, iPhone, and Meta Neural Band.}
    \label{fig:devices}
    \vspace{-4mm}
\end{figure}

\paragraph{Runtimes for different devices.}
To realize synthesized interfaces on heterogeneous hardware, we implement runtimes for Apple Vision Pro, Meta Quest 3, and Ray-Ban Meta glasses (\autoref{fig:devices}), covering both spatial and screen-space proxy interfaces. 
Across these platforms, we support three input families: spatial gesture input, touchscreen input, and discrete gesture input. 
Spatial gesture input is supported on Vision Pro and Quest 3, while touchscreen input is provided through an iPhone companion app connected via a cloud server and discrete gesture input comes from Meta's wristband. 
This prototype demonstrates feasibility of the capability class, although it does not yet match the fluidity of a dedicated commercial wristband.
Overall, our implementation is intended to demonstrate feasibility of the synthesis abstraction rather than optimize for any single device.

\final{\paragraph{System Response Time}
Because only the \texttt{SceneSpec} is acquired when proxy generation is invoked,
the end-to-end response time therefore comprises three components: (a) runtime scene-spec acquisition from the egocentric view, (b) candidate synthesis, and (c) candidate ranking.
We benchmarked the pipeline using the two scenes from our user study (Sec.~\ref{sec:studyprocedure}). The average end-to-end latency was below 6~s for both scenes: 4.22$\pm$0.31s for Bookshelf and 5.55$\pm$0.23s for Makerspace. The deterministic candidate-synthesis step (b) was lightweight, with latency dominated by the ML/LLM calls in steps (a) and (c). These results suggest that the current prototype is sufficiently responsive for on-demand proxy generation.
}

\vspace{-4mm}
\section{Application Scenarios}

We use three scenarios to probe two aspects of our contribution. 
First, they illustrate that proxy-based UIs can support rich interaction with real-world objects across a broader range of AR devices than direct target interaction alone. 
Second, they show that \CoolName{} can generate different proxy UIs for the same scene and intent under different device capability constraints. 
\final{The novelty illustrated by these scenarios lies not in individual operations such as filtering, semantic grouping, or hierarchical navigation, but in synthesizing proxy UIs that enable these operations across heterogeneous AR capability profiles.}
Accordingly, we present the scenarios as design probes rather than polished end-user applications. Together, they span dense object disambiguation on lightweight glasses, semantic regrouping for robot-mediated interaction, and multiscale navigation in building-scale environments.

\begin{figure}[h]
    \centering
    \includegraphics[width=1\linewidth]{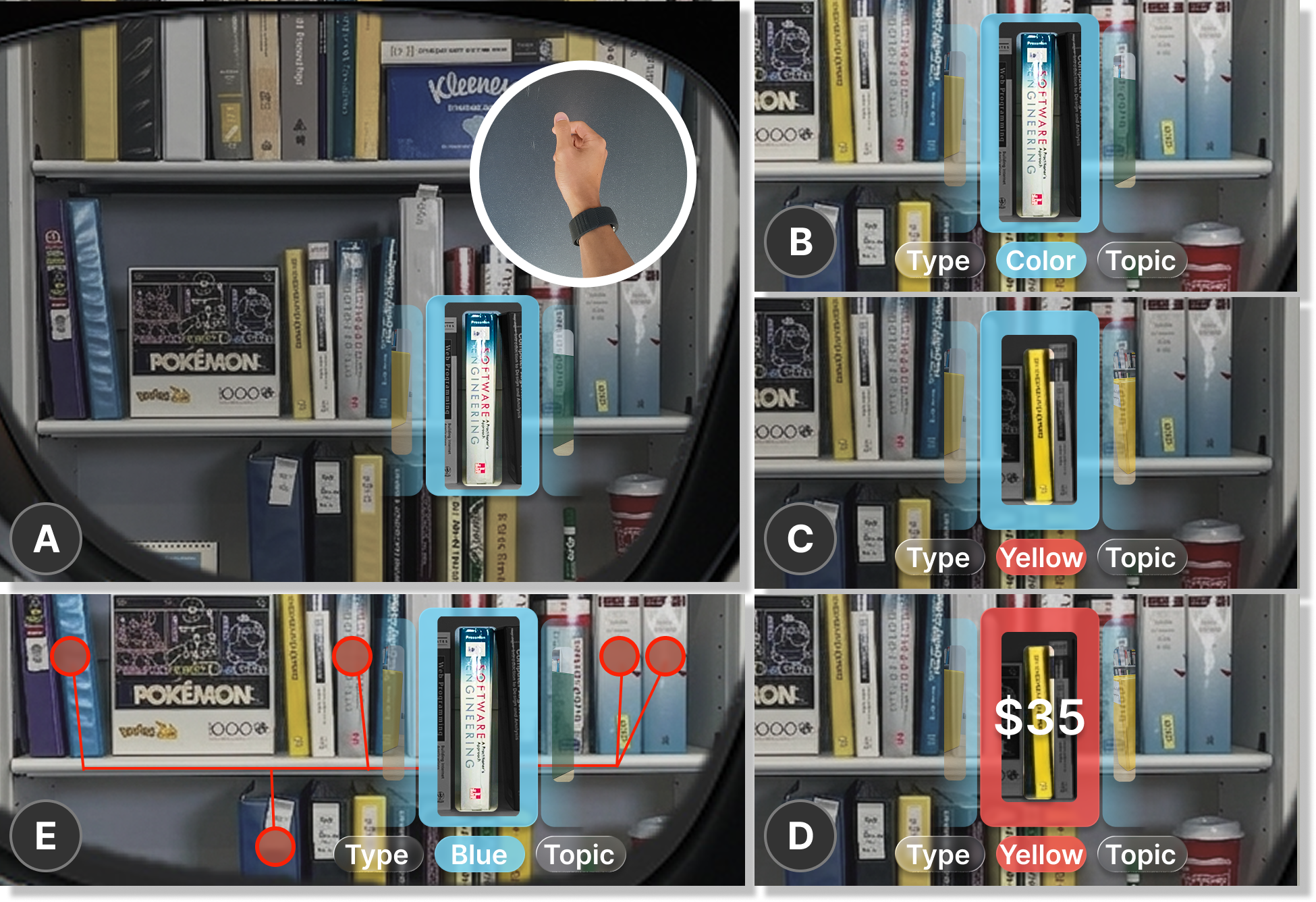}
       \vspace{-6mm}
    \caption{Scenario 1 — retrieving bookshelf item information using generated \final{symbolic Proxy UIs, corresponding to discrete input and visual 2D output. A-D show the interface on a partial 2D display, while E shows the interface on a full 2D display, where leader lines connect interface elements to their corresponding physical objects.}}
    \label{fig:Scene1Everyday}
    \vspace{-5mm}
\end{figure}

\subsection{Scenario 1: Everyday Information Retrieval}
\label{sec:bookshelf}

\noindent
This scenario examines dense everyday scenes in which direct target selection is difficult.
Suppose a user wants to retrieve the price of a yellow-covered book. 
On lightweight glasses, users may lack ray casting or eye tracking to precisely select one item from many visually similar books on a shelf. 
The challenge is therefore not only information retrieval, but also referential disambiguation.

Proxy-based UIs address this by replacing direct target selection with a generated representation over the visible objects. 
Instead of selecting a specific book in place, the user interacts with a compact proxy list that can be traversed using left--right gestures (\autoref{fig:Scene1Everyday}A). 
The user can then open an attribute filter such as \textit{Type}, \textit{Color}, or \textit{Topic} (\autoref{fig:Scene1Everyday}B). 
Selecting \textit{Yellow} reduces the candidate set to matching items (\autoref{fig:Scene1Everyday}C), after which the user can inspect the remaining items and retrieve the price of the target book (\autoref{fig:Scene1Everyday}D). \final{Under the same scene and intent, \CoolName{} can generate proxy UIs with spatial links for full display AR glasses (\autoref{fig:Scene1Everyday}E).}

This scenario illustrates the proxy-UI claim that dense object selection can be redirected into a more manageable proxy space on devices with limited display area and low-precision input. 
At the system level, it shows that \CoolName{} can generate a compact, filterable proxy view tailored to those capability constraints.

\begin{figure}[h]
    \centering
    \includegraphics[width=1\linewidth]{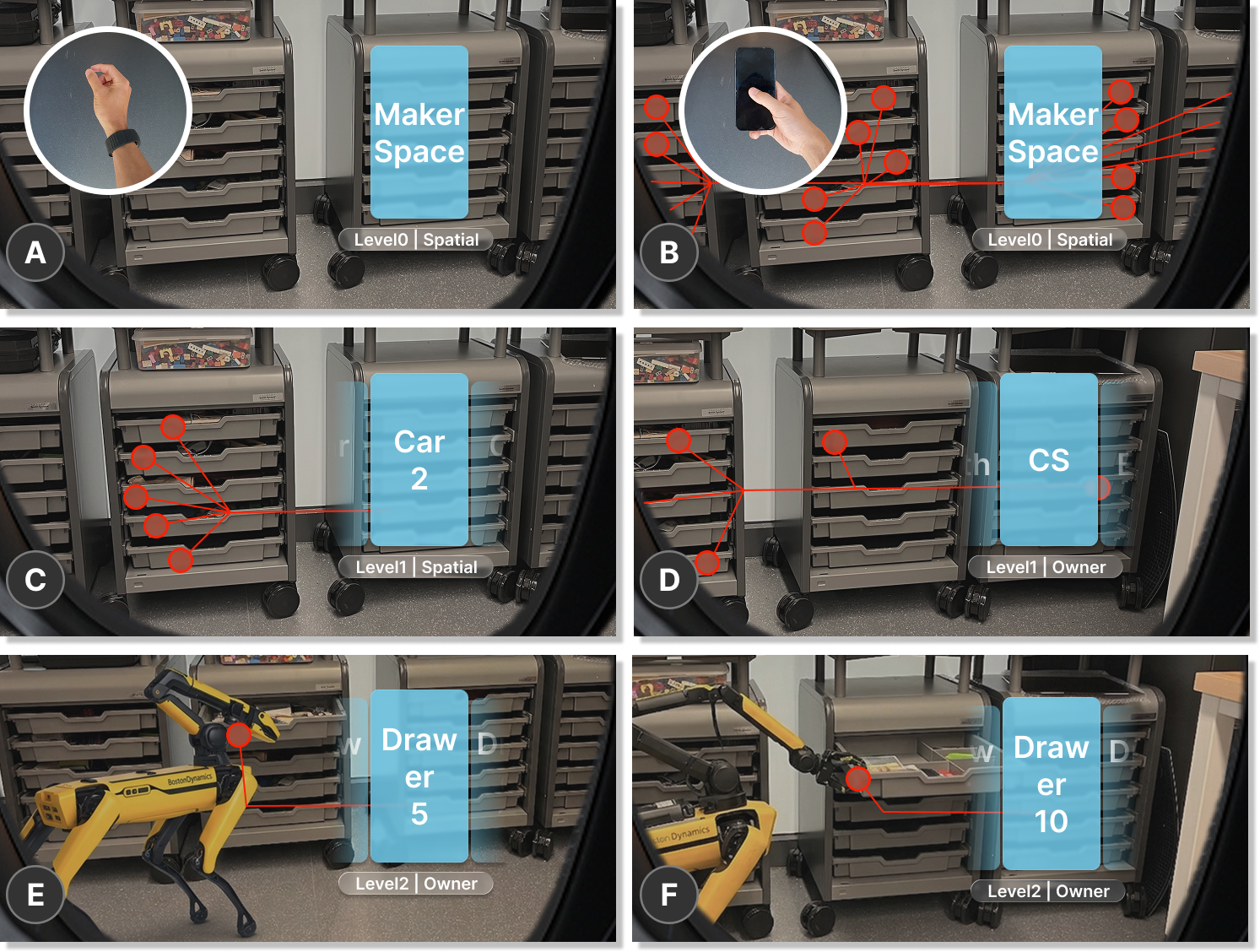}
       \vspace{-6mm}
    \caption{Scenario 2 — human--robot interaction in a makerspace using generated Proxy UIs. \final{Panel A shows symbolic UIs corresponding to discrete input and partial 2D output, while Panel B--F show 2D screen-pointer UIs corresponding to 2D pointer input and full 2D output.}}
    \label{fig:Scene2HRI}
    \vspace{-6mm}
\end{figure}

\subsection{Scenario 2: Human-Robot Collaboration}
\label{sec:robot}

\noindent
This scenario examines a different challenge: the physical organization of the environment may not match the semantic structure needed for the task.
In a maker space, a user wants to collaborate with a robot to retrieve materials from storage drawers based on semantic groupings such as ownership. 
However, the drawers are physically distributed across carts and shelves, making direct reference through speech or pointing awkward and difficult.

Proxy-based UIs address this by generating alternative structures over the same scene, intent, and across devices (\autoref{fig:Scene2HRI}A--B). 
\final{Rather than preserving only the original spatial hierarchy}, the system can regroup drawers by task-relevant attributes such as ownership. This allows the user to operate on the semantic abstraction needed for the task, while the robot maps the selected proxy back to the corresponding physical drawers. For example, in \autoref{fig:Scene2HRI}C, the default proxy reflects the original spatial organization of the carts. 
The user then switches to an ownership-based organization (\autoref{fig:Scene2HRI}D), which surfaces groups such as \textit{CS} and highlights the associated drawers. 
After selecting the \textit{CS} group, the user issues a command to request specific materials, and the robot retrieves items from the corresponding drawers (\autoref{fig:Scene2HRI}E--F).

This scenario illustrates that proxy-based UIs support more than easier selection under sensing or ergonomic constraints: they can also expose task-relevant semantic structure that is absent from the physical arrangement itself. 
For \CoolName{}, it highlights the ability to favor embedded feedback that preserves the link between proxies and physical referents when device capabilities permit.

\subsection{Scenario 3: Building Facility Management}
\label{sec:building}

This scenario examines large-scale environments in which direct interaction with individual physical referents is impractical. 
Building management tasks often require users to reason across multiple levels of structure, such as areas, sides, and rooms, and to reorganize that structure by semantic state rather than by physical layout alone. 
A direct one-object-at-a-time interaction model does not scale well to this setting.

\begin{figure}[h]
    \centering
    \includegraphics[width=1\linewidth]{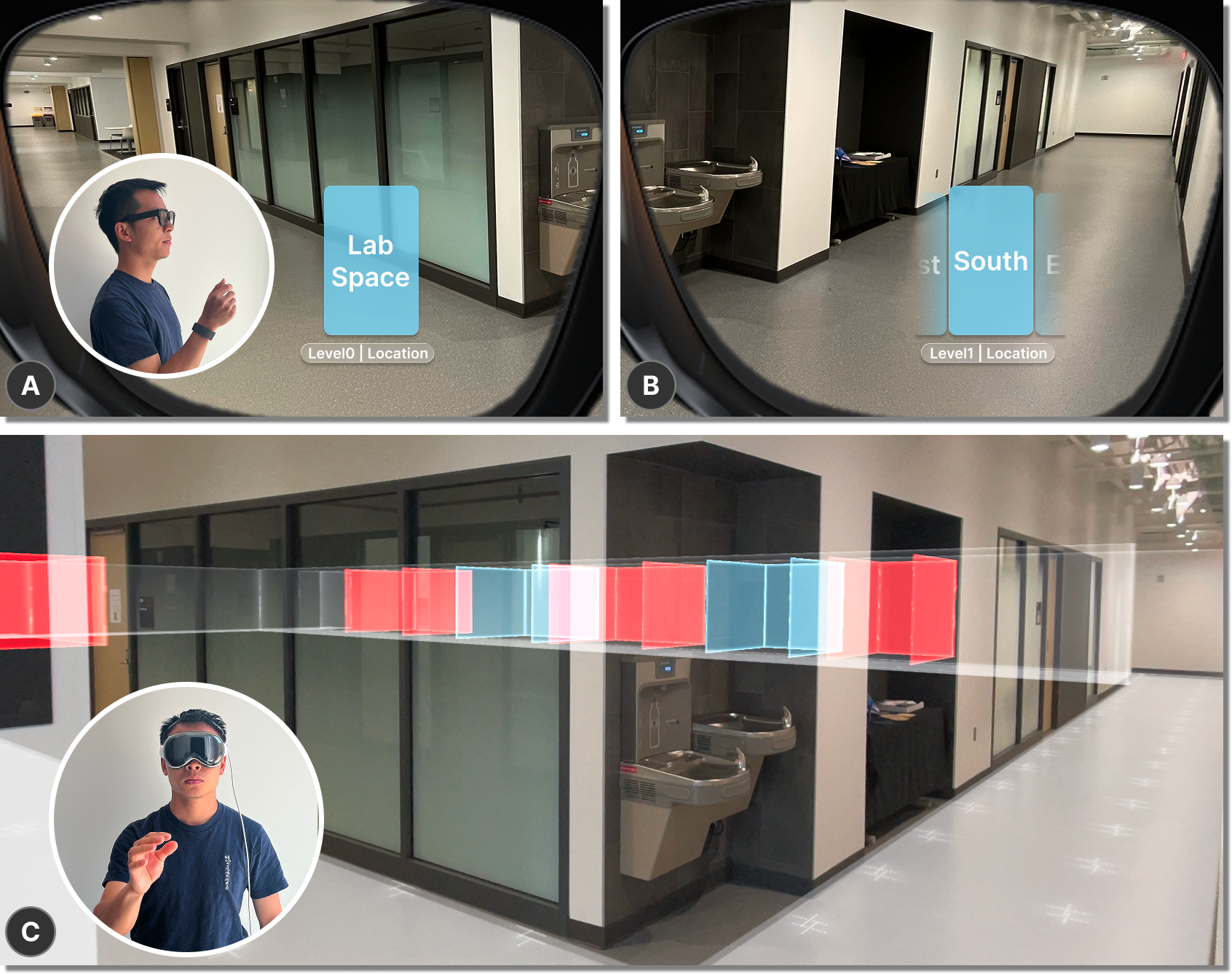}
       \vspace{-6mm}
    \caption{Scenario 3 — indoor building facility management using generated Proxy UIs. \final{Panels A--B show symbolic UIs corresponding to discrete gesture input and partial 2D output, while panel C shows spatial 3D UIs corresponding to spatial gesture input and spatial 3D output.}}
    \vspace{-6mm}
    \label{fig:Scene3Indoor}
\end{figure}

Proxy-based UIs address this by generating hierarchical proxies that support both multiscale navigation and semantic re-leveling. 
For example, on partial-display glasses, the user navigates a compact location hierarchy from the lab space to a particular side of the floor (\autoref{fig:Scene3Indoor}A--B).
Because proxies are generated from scene, intent, and capability specifications, the same underlying scene and task can be realized differently on different devices. 
Under a more immersive capability profile, \CoolName{} generates a spatially embedded proxy view of the same building structure (\autoref{fig:Scene3Indoor}C), making room-level relationships visible at a glance rather than through sequential hierarchical traversal.

More broadly, this scenario shows that proxy-based UIs can scale from local object collections to building-scale scenes, while \CoolName{} adapts their form to the target device's capabilities.

\section{Internal Ranking Sanity Check}

We conducted an internal blinded ranking analysis as a sanity check on the ranking pipeline. 
\final{Our goal was to assess whether the pipeline generally surfaced sensible candidates and to identify where its judgments broke down, rather than to provide definitive validation or broad comparative evaluation.
No single neutral baseline applies across tasks and device profiles, since hand-designed and rule-based alternatives embed their own design choices.}
We therefore report success rates and representative failure cases, leaving controlled, task-specific comparisons to future work.

We constructed a factorized benchmark of 50 cases spanning
2 scene types \final{(\emph{art gallery} and \emph{widget hierarchy})},
5 device profiles \final{(\emph{I:discrete gesture, O:partial visual 2D};
\emph{I:pointer 2D, O:full visual 2D};
\emph{I:discrete gesture, O:spatial 3D};
\emph{I:spatial 3D, O:full visual 2D};
\emph{I:spatial 3D, O:spatial 3D})}, and 5 intent bundles
\final{(\emph{inspect+select},
\emph{inspect+select+single filter},
\emph{inspect+select+double filter},
\emph{inspect+select+\\organize+relevel}, and
\emph{all intents})}. For each case, we synthesized all viable interface candidates, ranked them with our method, and sampled three candidates from different parts of the ranked list: top, middle, and lower. 
These candidates were anonymized and presented together with the corresponding scene description, device capability profile, and ordered task intents.
Two authors who were not involved in building the synthesizer or ranker independently ranked the three candidates, represented as UI specifications, from best to worst according to how well they supported the intended task. 
We evaluated specifications rather than deployed AR interfaces because reviewing all $50 \times 3 = 150$ candidates in-headset would have been fatiguing and impractical. 
We then compared the manual rankings with the system ranking using top-pick and full-order agreement; \final{a representative packet appears in ~\autoref{sec:ranking-candidates}.}

\paragraph{Results.}
For Coder 1, the system matched the top-ranked candidate in 98\% of cases and the full ordering in 66\%. 
For Coder 2, top-pick agreement was 76\% and full-order agreement was 62\%. 
Inter-rater agreement between the coders was $\kappa = 0.64$, indicating substantial consistency for this diagnostic analysis.

We examined cases in which the system's top choice did not match the coders' top choice and found three recurring causes. 
First, \textbf{preference divergence in interaction decomposition} (7/14): the system preferred interfaces that merged \emph{select} (or \emph{focus}) and \emph{inspect} into a single action to reduce articulatory distance, whereas coders often preferred separating these steps. 
Second, \textbf{ambiguity from gesture reuse across representations} (3/14): the same gesture was reused across different representations or states; although valid in the specification, coders judged these mappings as potentially ambiguous, especially without access to the realized interface. 
Third, \textbf{near-ties and residual evaluator disagreement} (4/14): the remaining mismatches appeared to reflect near-ties, subjective preference, or occasional evaluator oversight rather than a systematic ranking failure.

Overall, this analysis suggests that the current pipeline is effective at surfacing strong top candidates while also revealing actionable opportunities for improvement, particularly in modeling preferences around interaction decomposition and penalizing potentially ambiguous gesture reuse.







\section{Expert Evaluation}
We position \CoolName{} as a proof-of-concept and technology probe, aimed at assessing whether generated proxy UIs are plausible and useful under varied AR capability conditions.
To assess the perceived usefulness and limitations of the generated interfaces, we conducted an expert evaluation similar to prior technology-probe and expert-review studies in HCI~\cite{han2020textlets, liu2025reality}. \final{To compare with existing speech-based reference interaction ~\cite{lee2018interaction} on lightweight AR glasses, we also conducted a supplemental study.}

\subsection{Participants and Apparatus}
\label{sec:participants}
The study procedure and materials received IRB approval. 
We recruited 14 experienced XR developers and researchers from a local university's VR clubs and research groups. 
Two participants joined pilot sessions used to refine the protocol and were excluded from the reported ratings, leaving 12 participants for analysis. 6 of them had more than five years of AR/VR experience, 3 had more than two years, and 11 rated their headset proficiency above 8 on a 10-point scale. 
Each participant received \$30 for a 90-minute session.

The study was conducted in a controlled indoor environment using a Meta Quest 3 running a Unity application. 
Although we implemented multiple device-specific runtimes, we used Quest 3 as a common study platform to simulate the target input/output conditions. 
We chose this because our Apple Watch IMU prototype was sufficient to demonstrate feasibility but did not provide interaction fluidity comparable to a dedicated wrist-worn input device; using it directly would therefore have confounded the discrete gesture condition with prototype limitations rather than the interaction concept itself. 
To simulate AR glasses conditions, we constrained the Quest 3 interface to the same effective capability profile, including screen-space rendering, matched display size, and the corresponding input mappings. 
Thus, all study conditions shared the same synthesis backend and application logic, while differing only in the realized capability condition. 
All sessions were recorded in audio and video, including both first- and third-person perspectives, for post-study analysis.

\begin{figure}[h]
    \centering
    \includegraphics[width=1\linewidth]{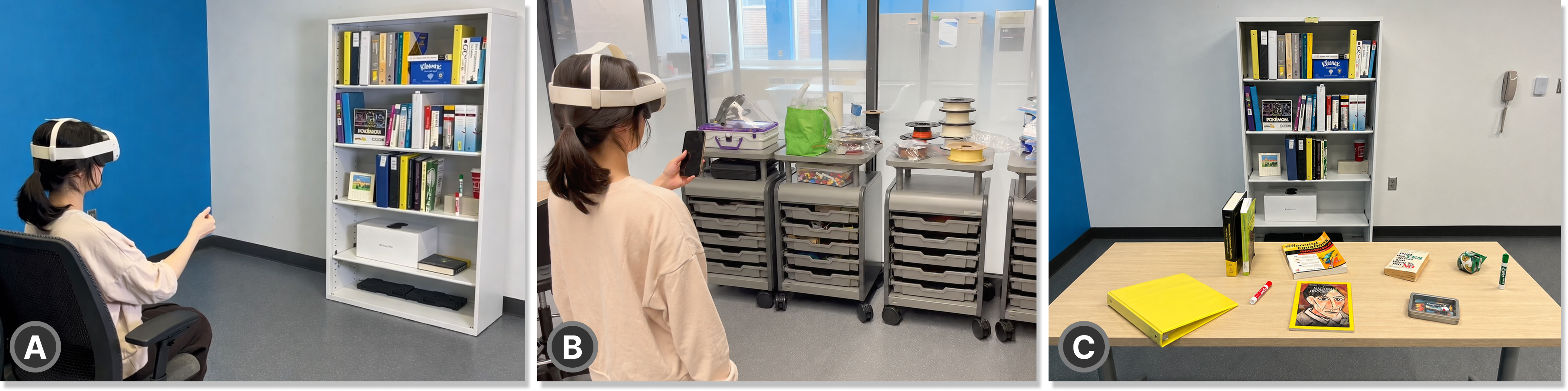}
    \vspace{-6mm}
    \caption{(A) Participant completing the bookshelf item-selection task. 
    (B) Participant completing the makerspace material-finding task. 
    (C) Supplemental study scene featuring a near-field tabletop and a far-field bookshelf.}
    \label{fig:study}
    \vspace{-6mm}
\end{figure}

\begin{figure*}[b]
  \centering
  \includegraphics[width=1\textwidth]{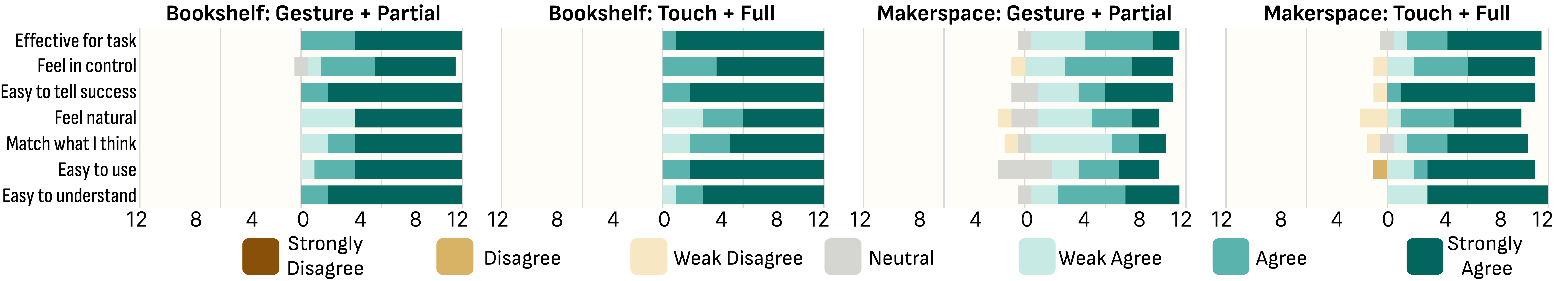}
  \caption{User feedback on UIs generated by \CoolName{} across different scene, intent, and device specifications.}
  \label{fig:userfeedback}
\end{figure*}

\subsection{Study Procedure}
\label{sec:studyprocedure}

Each session lasted approximately 90 minutes and consisted of training, task trials, and post-task feedback. 
During training, participants were introduced to the goals and the proxy-UI interaction concepts, then completed a guided tutorial using a simplified example until they felt comfortable with the interaction techniques.

Participants then completed two tasks: 
(1) \emph{Bookshelf Item Selection}, adapted from the bookshelf scenario (\autoref{sec:bookshelf}), in which they filtered shelf items and selected items under a budget constraint; and (2) \emph{Maker Space Material Finding}, adapted from the maker-space scenario (\autoref{sec:robot}), in which they identified drawers belonging to a designated cart and drawers associated with a target group. 
Each participant performed both tasks under two I/O conditions: 
(a) partial 2D display (simulated Meta Ray Ban Display) with micro-gesture input and 
(b) full 2D display with mobile touchscreen input. 
This yielded four trials per participant ($2$ tasks $\times$ $2$ I/O conditions), with order counterbalanced using a Latin square.

After each trial, participants completed a 7-point Likert questionnaire assessing the UI. 
They also ranked the UI with another two candidates sampled in advance from the middle and bottom of the system ranking for the corresponding scene and device-capability condition. 
We then conducted a semi-structured interview to collect qualitative feedback on the usefulness, limitations, and potential application scenarios of \CoolName{}.

\subsection{Expert Feedback}

\autoref{fig:userfeedback} shows that experts rated the generated proxy UIs positively across both tasks and both device conditions. 
To understand the reasons behind these ratings, we analyzed the interview data using reflexive thematic analysis~\cite{braun2019reflecting}. 
Two authors collaboratively coded the interview transcripts and grouped the resulting codes into three themes.
Unless otherwise noted, the ratings discussed below are averaged across the four study cases.

\subsubsection{Overall usability and utility of proxy-based UIs on constrained AR devices}
Experts generally found proxy-based UIs easy to understand (6.5/7), easy to use (6.25/7), controllable (6.17/7), and effective across tasks (6.40/7). 
\final{These results suggest that proxy interaction remained usable under the simulated lightweight-glasses capability constraints. 
Participants particularly valued compact proxy representations when limited display area and low-precision input made spatial target selection cumbersome (P3, P9). Rather than pointing precisely at distant or crowded objects, users could navigate, filter, and select within a compact interface. Attribute filters narrowed visually similar items in the bookshelf task (P4, P11), while hierarchical grouping exposed task-relevant structure in the maker-space task (P5). Overall, participants regarded proxy UIs as a practical way to support real-world object interaction when device capabilities do not support precise spatial targeting.
}

\subsubsection{Feedback on generated proxy UIs across device configurations}

Experts also commented on whether the generated proxy UIs appropriately matched different device conditions. 
Across conditions, ratings for \emph{easy to tell success} remained consistently positive (6.52/7), suggesting that the generated UIs generally made action outcomes legible. 
Ratings for \emph{feel natural} (5.92/7) and \emph{match what I think} (6.04/7) were also positive overall, but varied more across conditions, indicating that these dimensions were more sensitive to the fit between proxy form and device capability.

\begin{table}[th]
\centering
\caption{Agreement between system and expert rankings. Top Pick indicates agreement on the best candidate; Full indicates exact agreement on the full three-way ranking.}
\vspace{-4mm}
\label{tab:hit-rate-results}
\begin{tabular}{l l c c}
\toprule
Scenario & Condition & Top Pick & Full \\
\midrule
Bookshelf   & Partial + Gesture & 83.33\% & 58.33\% \\
Bookshelf   & Full + Touch      & 91.67\% & 58.33\% \\
Maker Space & Partial + Gesture & 91.67\% & 58.33\% \\
Maker Space & Full + Touch      & 100\%   & 100\%   \\
\bottomrule
\end{tabular}
\vspace{-4mm}
\end{table}

We also compared the system's ranking of candidate UIs with expert judgments. 
As shown in \autoref{tab:hit-rate-results}, the system's top-ranked candidate matched the experts' top choice in 83.33\% to 100\% of cases, while full ranking agreement ranged from 58.33\% to 100\%. 
Agreement was strongest in the \emph{Full + Touch} makerspace condition, where the system exactly matched expert preferences. 
Overall, these results suggest that the ranker captures expert preference reasonably well, especially when identifying the best candidate, whereas differences between middle- and lower-ranked alternatives were less distinct.

Qualitative feedback helps explain this pattern. 
When device capabilities allowed stronger visual linkage to the physical environment, participants preferred proxies with linked highlights, dots, and lines, which made referents easier to identify and action outcomes easier to interpret (P4, P5, P8). P4 noted, \quot{Linking the proxy to real-world objects really helps convey the association}.
In more constrained partial-display conditions, participants (\eg, P7) still found the generated proxies useful, but relied more on compact organization, filtering, and hierarchical navigation when direct visual correspondence was limited.
This helps explain why evaluation-related ratings remained strong across conditions, while naturalness and semantic fit were more mixed in the constrained cases.

Experts also noted that different device conditions called for different proxy forms. 
Compact, filterable views were seen as appropriate for \emph{Gesture + Partial} interaction (P2, P3, P9, P12), where display space and input bandwidth were limited, whereas \emph{Touch + Full} conditions benefited from richer spatial correspondence and more persistent links to scene objects (P2, P6, P8).
At the same time, participants (P6, P7, P11) appreciated that core operations such as traversing, filtering, and selecting remained conceptually stable across configurations. 
Together, feedback supports the core premise of \CoolName{}: generated proxy UIs should adapt to device capability while preserving a consistent interaction vocabulary.

\subsubsection{Design opportunities and limitations}
Despite the generally positive feedback, experts also identified several opportunities for improvement. First, while hierarchy and filtering were useful, participants (\eg, P12) noted that proxy structures could become harder to read in constrained partial-display conditions. In particular, the tree structure was not always immediately legible when only a small portion of the interface was visible at once. Commented by P2, \quot{Visible tree structure would help map my mental model with the physical model.} This suggests a need for stronger visual support for hierarchy, such as clearer level indicators, progressive disclosure, or lightweight overview cues.
Second, participants emphasized that feedback should remain as tightly coupled to the physical scene as possible. When device capabilities supported embedded visual correspondence, experts preferred in-scene or scene-linked feedback because it made action outcomes easier to interpret. P6 highlighted that \quot{I definitely liked the lines going to the drawers. It helped me really understand what the hierarchy is showing me.}
When such feedback was absent or reduced, the interaction remained usable, but felt more abstract and required more cognitive effort (P11). This reinforces the importance of evaluation legibility in generated proxy UIs.

\subsection{Supplemental Study}
\final{
To examine when proxy-based interaction is preferable to direct reference on lightweight AR glasses, we conducted a supplemental within-subject study comparing proxy UIs generated by \CoolName{} with direct speech-based reference.
We used speech as a representative direct-reference modality that allows users to identify real-world targets without first navigating an intermediate interface~\cite{lee2018interaction}.
Seven XR experts from the participant pool described in Sec.~\ref{sec:participants} evaluated both methods in two settings (\autoref{fig:study}C): a sparse, near-field tabletop scene (0.5m) and a dense, far-field bookshelf scene (3m).

Participants completed information retrieval tasks by filtering objects and retrieving prices encoded in their metadata.
In the speech condition, participants referred directly to target objects using speech; in the proxy condition, they navigated and selected from a generated proxy UI using the Meta Neural Band.
Both conditions used the Meta Ray-Ban Display.
Each session lasted approximately 30 mins, including training, tasks, and a semi-structured interview.

Six of seven participants preferred direct speech-based reference in the sparse tabletop setting, whereas six of seven preferred proxy-based interaction in the dense bookshelf setting.
Participants explained that speech was efficient when objects were clearly visible and easy to name, and that it offered flexibility for open-ended requests (E2, E3).
In contrast, they preferred proxy UIs when objects were crowded, difficult to distinguish, or difficult to describe precisely, because the proxy provided explicit alternatives for disambiguation (E3, E7).

Because the two settings jointly varied distance, density, and target nameability, these results do not isolate the effect of each factor independently.
Instead, they provide initial evidence that proxy-based interaction complements direct speech-based reference on lightweight AR glasses: speech may be more natural when targets are readily perceivable and nameable, whereas proxies may be more useful when direct reference is ambiguous or cumbersome.

}

\section{Discussion}

\subsection{Proxy UIs as a Capability-bridging Interaction Layer}
Our results suggest that the key value of proxy UIs is not only that they avoid difficult direct targeting, but that they introduce a digital interaction layer that is more fluid than the physical scene itself. 
Proxies can be reorganized, regrouped, and re-leveled at runtime to match the user's task and the capabilities of the current device. 
This makes them a capability-bridging layer: they preserve the semantics of real objects while freeing interaction from the fixed layout of the environment and the limitations of any single device.

This perspective helps explain why proxy form should vary across platforms. 
Richer devices can support tighter scene linkage and embedded feedback, while more constrained devices benefit from compact panels, hierarchical navigation, or symbolic interaction. 
More broadly, proxy-based UI should be understood not as a workaround for limited AR hardware, but as a general strategy for transforming real-world interaction into adaptive digital representations that remain grounded in the physical world.

\subsection{Generating AR UIs as Miniatures of the User's Mental Model}

Although we do not formulate \CoolName{} as a generative model in the probabilistic sense, the interface is synthesized at runtime from the current scene, user intent, and device capability, and LLM-based judgments in the ranking stage. 
In this sense, the system does not retrieve a fixed interface template, but generates a proxy UI specific to the current situation.

More importantly, our results suggest that effective generative UI in AR is not only a matter of producing functionally correct widgets or layouts. 
Because the interface mediates interaction with the physical world, it must also reflect how users mentally organize that world for the task at hand. 
In our scenarios, the most useful proxies were those that exposed task-relevant structure already implicit in the environment, such as grouping books by topic, drawers by ownership, or spaces by hierarchical location. 
These proxies function as interactive miniatures of the user's working mental model: compact representations that preserve the distinctions most relevant for action on a given device.

This suggests that generative UI in AR should be understood not simply as interface generation, but as the synthesis of interactive representations that align the user's mental model, the physical scene, and the device's capabilities. 
A key next step is therefore richer user modeling, so that generated proxies can better reflect differences in users' goals, prior knowledge, and preferred ways of structuring the world.

\vspace{-2mm}
\subsection{Limitations and Future Work}
\label{sec:futurework}
Our study used Meta Quest 3 as a controlled simulator for multiple capability conditions.
Future work should validate the system on a broader set of native deployments.
\final{Moreover, the current latency of proxy UI generation is compatible with the intended use of \CoolName{} for user-triggered, on-demand synthesis rather than continuous per-frame generation. Future work could further reduce latency through caching and parallelization, as well as by replacing cloud-based LLM calls with faster local or task-specific models.}
Our expert evaluation focused on experienced XR users, most of whom were university-affiliated. Broader studies are needed to understand how proxy-based UIs generalize to less technical users and to longer-term everyday use. Proxy-based UI is better understood as an interaction paradigm than as one low-level technique, and its appropriate comparison point may vary across tasks and devices. Future work should compare generated proxies against task-specific direct interaction baselines where appropriate.
Finally, our current implementation focuses on visual and gesture-based interaction and does not yet explore voice as an input modality. Voice may be especially valuable for lightweight AR devices and could complement proxy navigation in future systems.
\section{Conclusions}

We introduced \CoolName{}, a method for synthesizing proxy-based interfaces for interacting with real-world objects across heterogeneous AR capability conditions. Grounded in constrained synthesis and a ranking procedure based on semantic, articulatory, and evaluation distance, \CoolName{} takes structured specifications of the scene, the user's intent, and the target device to generate proxy-based interfaces. By decoupling interaction from the physical constraints of the environment and the limitations of any single device, these interfaces enable users to interact more effectively with objects and spaces that are dense, distant, cluttered, or otherwise difficult to access through direct AR interaction. Through several application scenarios, we demonstrated the versatility of this approach across diverse scenes, intents, and AR form factors. Our internal ranking sanity check and expert evaluation further highlight the promise of generated proxy UIs as a capability-bridging interaction layer for AR, suggesting a path toward more fluid, adaptive, and expressive interaction with real-world objects across the next generation of AR glasses.

\begin{acks}
We thank the reviewers for their constructive feedback. 
This work is partially supported by NSF 2543054 and Google.
\end{acks}

\bibliographystyle{ACM-Reference-Format}
\bibliography{reference}

@String{Computing = "Computing" }

@String{Computer = "{IEEE} Computer" }

@String{Springer = "Springer-Verlag" }

@inproceedings{liu2025reality,
  title={Reality proxy: fluid interactions with real-world objects in MR via abstract representations},
  author={Liu, Xiaoan and Jia, Difan and Liu, Xianhao Carton and Gonzalez-Franco, Mar and Zhu-Tian, Chen},
  booktitle={Proceedings of the 38th Annual ACM Symposium on User Interface Software and Technology},
  pages={1--16},
  year={2025}
}

@inproceedings{han2020textlets,
  title={Textlets: Supporting constraints and consistency in text documents},
  author={Han, Han L and Renom, Miguel A and Mackay, Wendy E and Beaudouin-Lafon, Michel},
  booktitle={Proceedings of the 2020 CHI Conference on Human Factors in Computing Systems},
  pages={1--13},
  year={2020}
}

@article{pfeuffer2024design,
  title={Design principles and challenges for gaze+ pinch interaction in xr},
  author={Pfeuffer, Ken and Gellersen, Hans and Gonzalez-Franco, Mar},
  journal={IEEE Computer Graphics and Applications},
  volume={44},
  number={03},
  pages={74--81},
  year={2024},
  publisher={IEEE Computer Society}
}

@article{gonzalez2024guidelines,
  title={Guidelines for productivity in virtual reality},
  author={Gonzalez-Franco, Mar and Colaco, Andrea},
  journal={Interactions},
  volume={31},
  number={3},
  pages={46--53},
  year={2024},
  publisher={ACM New York, NY, USA}
}

@inproceedings{suzuki2025everyday,
  title={Everyday AR through AI-in-the-Loop},
  author={Suzuki, Ryo and Gonzalez-Franco, Mar and Sra, Misha and Lindlbauer, David},
  booktitle={Proceedings of the Extended Abstracts of the CHI Conference on Human Factors in Computing Systems},
  pages={1--5},
  year={2025}
}

@inproceedings{lee2020ubipoint,
  title={UbiPoint: towards non-intrusive mid-air interaction for hardware constrained smart glasses},
  author={Lee, Lik Hang and Braud, Tristan and Bijarbooneh, Farshid Hassani and Hui, Pan},
  booktitle={Proceedings of the 11th ACM Multimedia Systems Conference},
  pages={190--201},
  year={2020}
}

@article{nguyen2023hand,
  title={Hand interaction designs in mixed and augmented reality head mounted display: a scoping review and classification},
  author={Nguyen, Richard and Gouin-Vallerand, Charles and Amiri, Maryam},
  journal={Frontiers in Virtual Reality},
  volume={4},
  pages={1171230},
  year={2023},
  publisher={Frontiers Media SA}
}

@inproceedings{jiang2019orc,
  title={ORC layout: Adaptive GUI layout with OR-constraints},
  author={Jiang, Yue and Du, Ruofei and Lutteroth, Christof and Stuerzlinger, Wolfgang},
  booktitle={Proceedings of the 2019 CHI Conference on human factors in computing systems},
  pages={1--12},
  year={2019}
}

@inproceedings{swearngin2020scout,
  title={Scout: Rapid exploration of interface layout alternatives through high-level design constraints},
  author={Swearngin, Amanda and Wang, Chenglong and Oleson, Alannah and Fogarty, James and Ko, Amy J},
  booktitle={Proceedings of the 2020 CHI conference on human factors in computing systems},
  pages={1--13},
  year={2020}
}

@inproceedings{zanden1990automatic,
  title={Automatic, look-and-feel independent dialog creation for graphical user interfaces},
  author={Zanden, Brad Vander and Myers, Brad A},
  booktitle={Proceedings of the SIGCHI conference on Human factors in computing systems},
  pages={27--34},
  year={1990}
}

@inproceedings{szekely1993beyond,
  title={Beyond interface builders: Model-based interface tools},
  author={Szekely, Pedro and Luo, Ping and Neches, Robert},
  booktitle={Proceedings of the INTERACT'93 and CHI'93 Conference on Human Factors in Computing Systems},
  pages={383--390},
  year={1993}
}

@inproceedings{nichols2004improving,
  title={Improving automatic interface generation with smart templates},
  author={Nichols, Jeffrey and Myers, Brad A and Litwack, Kevin},
  booktitle={Proceedings of the 9th international conference on Intelligent user interfaces},
  pages={286--288},
  year={2004}
}

@inproceedings{miniotas2006speech,
  title={Speech-augmented eye gaze interaction with small closely spaced targets},
  author={Miniotas, Darius and {\v{S}}pakov, Oleg and Tugoy, Ivan and MacKenzie, I Scott},
  booktitle={Proceedings of the 2006 symposium on Eye tracking research \& applications},
  pages={67--72},
  year={2006}
}

@inproceedings{chatterjee2015gaze,
  title={Gaze+ gesture: Expressive, precise and targeted free-space interactions},
  author={Chatterjee, Ishan and Xiao, Robert and Harrison, Chris},
  booktitle={Proceedings of the 2015 ACM on international conference on multimodal interaction},
  pages={131--138},
  year={2015}
}

@article{chen2021evaluating,
  title={Evaluating large language models trained on code},
  author={Chen, Mark and Tworek, Jerry and Jun, Heewoo and Yuan, Qiming and Pinto, Henrique Ponde De Oliveira and Kaplan, Jared and Edwards, Harri and Burda, Yuri and Joseph, Nicholas and Brockman, Greg and others},
  journal={arXiv preprint arXiv:2107.03374},
  year={2021}
}

@incollection{lee2023embodied,
  title={Embodied interaction on constrained interfaces for augmented reality},
  author={Lee, Lik-Hang and Braud, Tristan and Hui, Pan},
  booktitle={Springer Handbook of Augmented Reality},
  pages={239--271},
  year={2023},
  publisher={Springer}
}

@article{lai2024eye,
  title={In the eye of transformer: Global--local correlation for egocentric gaze estimation and beyond},
  author={Lai, Bolin and Liu, Miao and Ryan, Fiona and Rehg, James M},
  journal={International Journal of Computer Vision},
  volume={132},
  number={3},
  pages={854--871},
  year={2024},
  publisher={Springer}
}

@inproceedings{suzuki2020realitysketch,
  title={Realitysketch: Embedding responsive graphics and visualizations in AR through dynamic sketching},
  author={Suzuki, Ryo and Kazi, Rubaiat Habib and Wei, Li-yi and DiVerdi, Stephen and Li, Wilmot and Leithinger, Daniel},
  booktitle={Proceedings of the 33rd Annual ACM Symposium on User Interface Software and Technology},
  pages={166--181},
  year={2020}
}

@inproceedings{lee2024gazepointar,
  title={GazePointAR: A context-aware multimodal voice assistant for pronoun disambiguation in wearable augmented reality},
  author={Lee, Jaewook and Wang, Jun and Brown, Elizabeth and Chu, Liam and Rodriguez, Sebastian S and Froehlich, Jon E},
  booktitle={Proceedings of the 2024 CHI Conference on Human Factors in Computing Systems},
  pages={1--20},
  year={2024}
}

@article{tsai2026uncertain,
  title={Uncertain Pointer: Situated Feedforward Visualizations for Ambiguity-Aware AR Target Selection},
  author={Tsai, Ching-Yi and Tacconi, Nicole and Wilson, Andrew D and Abtahi, Parastoo},
  journal={arXiv preprint arXiv:2602.13433},
  year={2026}
}

@inproceedings{lee2025sensible,
  title={Sensible agent: A framework for unobtrusive interaction with proactive ar agents},
  author={Lee, Geonsun and Xia, Min and Numan, Nels and Qian, Xun and Li, David and Chen, Yanhe and Kulshrestha, Achin and Chatterjee, Ishan and Zhang, Yinda and Manocha, Dinesh and others},
  booktitle={Proceedings of the 38th Annual ACM Symposium on User Interface Software and Technology},
  pages={1--22},
  year={2025}
}

@inproceedings{fashimpaur2025squiggle,
  title={Squiggle: Multimodal Lasso Selection in the Real World},
  author={Fashimpaur, Jacqui and Grossman, Tovi and Lafreniere, Ben and Sendhilnathan, Naveen and Todi, Kashyap and Wang, Tianyi and Zhang, Ting and Jonker, Tanya R},
  booktitle={Proceedings of the 38th Annual ACM Symposium on User Interface Software and Technology},
  pages={1--16},
  year={2025}
}

@inproceedings{vaithilingam2019bespoke,
  title={Bespoke: Interactively synthesizing custom GUIs from command-line applications by demonstration},
  author={Vaithilingam, Priyan and Guo, Philip J},
  booktitle={Proceedings of the 32nd annual ACM symposium on user interface software and technology},
  pages={563--576},
  year={2019}
}

@article{satyanarayan2016vega,
  title={Vega-lite: A grammar of interactive graphics},
  author={Satyanarayan, Arvind and Moritz, Dominik and Wongsuphasawat, Kanit and Heer, Jeffrey},
  journal={IEEE transactions on visualization and computer graphics},
  volume={23},
  number={1},
  pages={341--350},
  year={2016},
  publisher={IEEE}
}

@inproceedings{vaithilingam2024dynavis,
  title={Dynavis: Dynamically synthesized ui widgets for visualization editing},
  author={Vaithilingam, Priyan and Glassman, Elena L and Inala, Jeevana Priya and Wang, Chenglong},
  booktitle={Proceedings of the 2024 CHI Conference on Human Factors in Computing Systems},
  pages={1--17},
  year={2024}
}

@inproceedings{wu2024uicoder,
  title={Uicoder: Finetuning large language models to generate user interface code through automated feedback},
  author={Wu, Jason and Schoop, Eldon and Leung, Alan and Barik, Titus and Bigham, Jeffrey P and Nichols, Jeffrey},
  booktitle={Proceedings of the 2024 Conference of the North American Chapter of the Association for Computational Linguistics: Human Language Technologies (Volume 1: Long Papers)},
  pages={7511--7525},
  year={2024}
}

@inproceedings{lu2025misty,
  title={Misty: Ui prototyping through interactive conceptual blending},
  author={Lu, Yuwen and Leung, Alan and Swearngin, Amanda and Nichols, Jeffrey and Barik, Titus},
  booktitle={Proceedings of the 2025 CHI Conference on Human Factors in Computing Systems},
  pages={1--17},
  year={2025}
}

@inproceedings{cao2025generative,
  title={Generative and malleable user interfaces with generative and evolving task-driven data model},
  author={Cao, Yining and Jiang, Peiling and Xia, Haijun},
  booktitle={Proceedings of the 2025 CHI Conference on Human Factors in Computing Systems},
  pages={1--20},
  year={2025}
}

@inproceedings{min2025malleable,
  title={Malleable overview-detail interfaces},
  author={Min, Bryan and Chen, Allen and Cao, Yining and Xia, Haijun},
  booktitle={Proceedings of the 2025 CHI Conference on Human Factors in Computing Systems},
  pages={1--25},
  year={2025}
}

@inproceedings{kato1999marker,
  title={Marker tracking and hmd calibration for a video-based augmented reality conferencing system},
  author={Kato, Hirokazu and Billinghurst, Mark},
  booktitle={Proceedings 2nd IEEE and ACM International Workshop on Augmented Reality (IWAR'99)},
  pages={85--94},
  year={1999},
  organization={IEEE}
}

@inproceedings{henrysson2005face,
  title={Face to face collaborative AR on mobile phones},
  author={Henrysson, Anders and Billinghurst, Mark and Ollila, Mark},
  booktitle={Fourth ieee and acm international symposium on mixed and augmented reality (ismar'05)},
  pages={80--89},
  year={2005},
  organization={IEEE}
}

@article{chatterjee2025flowring,
  title={FlowRing: Integrated Microgesture and Surface Interaction Ring for Versatile XR Input},
  author={Chatterjee, Ishan and Ding, Jiexin and Waghmare, Anandghan and Breda, Joseph and Deng, Yuquan and Liu, Bo and Wang, Yuntao and Patel, Shwetak},
  journal={Proceedings of the ACM on Human-Computer Interaction},
  volume={9},
  number={5},
  pages={1--28},
  year={2025},
  publisher={ACM New York, NY}
}

@article{wang2025computing,
  title={Computing with smart rings: A systematic literature review},
  author={Wang, Zeyu and Yu, Ruotong and Wang, Xiangyang and Ding, Jiexin and Tang, Jiankai and Fang, Jun and He, Zhe and Li, Zhuojun and R{\"o}ddiger, Tobias and Xu, Weiye and others},
  journal={Proceedings of the ACM on Interactive, Mobile, Wearable and Ubiquitous Technologies},
  volume={9},
  number={3},
  pages={1--54},
  year={2025},
  publisher={ACM New York, NY, USA}
}

@article{caramiaux2015understanding,
  title={Understanding gesture expressivity through muscle sensing},
  author={Caramiaux, Baptiste and Donnarumma, Marco and Tanaka, Atau},
  journal={ACM Transactions on Computer-Human Interaction (TOCHI)},
  volume={21},
  number={6},
  pages={1--26},
  year={2015},
  publisher={ACM New York, NY, USA}
}

@article{braun2019reflecting,
  title={Reflecting on reflexive thematic analysis},
  author={Braun, Virginia and Clarke, Victoria},
  journal={Qualitative research in sport, exercise and health},
  volume={11},
  number={4},
  pages={589--597},
  year={2019},
  publisher={Taylor \& Francis}
}

@article{milgram1994taxonomy,
  title={A taxonomy of mixed reality visual displays},
  author={Milgram, Paul and Kishino, Fumio},
  journal={IEICE TRANSACTIONS on Information and Systems},
  volume={77},
  number={12},
  pages={1321--1329},
  year={1994},
  publisher={The Institute of Electronics, Information and Communication Engineers}
}

@inproceedings{liu2026can,
  title={Can AR Embedded Visualizations Foster Appropriate Reliance on AI in Spatial Decision-Making? A Comparative Study of AR X-Ray vs. 2D Minimap},
  author={Liu, Xianhao Carton and Jia, Difan and Nie, Tongyu and Suma Rosenberg, Evan and Interrante, Victoria and Zhu-Tian, Chen},
  booktitle={Proceedings of the 2026 CHI Conference on Human Factors in Computing Systems},
  pages={1--15},
  year={2026}
}

@book{laviola20173d,
  title={3D user interfaces: theory and practice},
  author={LaViola Jr, Joseph J and Kruijff, Ernst and McMahan, Ryan P and Bowman, Doug and Poupyrev, Ivan P},
  year={2017},
  publisher={Addison-Wesley Professional}
}

@inproceedings{dogan2024augmented,
  title={Augmented object intelligence with xr-objects},
  author={Dogan, Mustafa Doga and Gonzalez, Eric J and Ahuja, Karan and Du, Ruofei and Cola{\c{c}}o, Andrea and Lee, Johnny and Gonzalez-Franco, Mar and Kim, David},
  booktitle={Proceedings of the 37th Annual ACM Symposium on User Interface Software and Technology},
  pages={1--15},
  year={2024}
}

@article{tong2022exploring,
  title={Exploring interactions with printed data visualizations in augmented reality},
  author={Tong, Wai and Zhu-Tian, Chen and Xia, Meng and Lo, Leo Yu-Ho and Yuan, Linping and Bach, Benjamin and Qu, Huamin},
  journal={IEEE transactions on visualization and computer graphics},
  volume={29},
  number={1},
  pages={418--428},
  year={2022},
  publisher={IEEE}
}

@inproceedings{kyto2018pinpointing,
  title={Pinpointing: Precise head-and eye-based target selection for augmented reality},
  author={Kyt{\"o}, Mikko and Ens, Barrett and Piumsomboon, Thammathip and Lee, Gun A and Billinghurst, Mark},
  booktitle={Proceedings of the 2018 CHI Conference on Human Factors in Computing Systems},
  pages={1--14},
  year={2018}
}

@inproceedings{du2022opportunistic,
  title={Opportunistic interfaces for augmented reality: Transforming everyday objects into tangible 6dof interfaces using ad hoc ui},
  author={Du, Ruofei and Olwal, Alex and Le Goc, Mathieu and Wu, Shengzhi and Tang, Danhang and Zhang, Yinda and Zhang, Jun and Tan, David Joseph and Tombari, Federico and Kim, David},
  booktitle={CHI Extended Abstracts},
  pages={1--4},
  year={2022}
}

@inproceedings{monteiro2023teachable,
  title={Teachable reality: Prototyping tangible augmented reality with everyday objects by leveraging interactive machine teaching},
  author={Monteiro, Kyzyl and Vatsal, Ritik and Chulpongsatorn, Neil and Parnami, Aman and Suzuki, Ryo},
  booktitle={Proc. CHI},
  pages={1--15},
  year={2023}
}

@inproceedings{hettiarachchi2016annexing,
  title={Annexing reality: Enabling opportunistic use of everyday objects as tangible proxies in augmented reality},
  author={Hettiarachchi, Anuruddha and Wigdor, Daniel},
  booktitle={Proceedings of the 2016 CHI Conference on Human Factors in Computing Systems},
  pages={1957--1967},
  year={2016}
}

@inproceedings{kan2009applying,
  title={Applying QR code in augmented reality applications},
  author={Kan, Tai-Wei and Teng, Chin-Hung and Chou, Wen-Shou},
  booktitle={Proceedings of the 8th international conference on virtual reality continuum and its applications in industry},
  pages={253--257},
  year={2009}
}

@inproceedings{ahuja2019lightanchors,
  title={Lightanchors: Appropriating point lights for spatially-anchored augmented reality interfaces},
  author={Ahuja, Karan and Pareddy, Sujeath and Xiao, Robert and Goel, Mayank and Harrison, Chris},
  booktitle={Proceedings of the 32nd Annual ACM Symposium on User Interface Software and Technology},
  pages={189--196},
  year={2019}
}

@inproceedings{leung2025squire,
  title={SQUIRE: Interactive UI Authoring via Slot QUery Intermediate REpresentations},
  author={Leung, Alan and Cheng, Ruijia and Wu, Jason and Nichols, Jeffrey and Barik, Titus},
  booktitle={Proceedings of the 38th Annual ACM Symposium on User Interface Software and Technology},
  pages={1--17},
  year={2025}
}

@article{grubert2016towards,
  title={Towards pervasive augmented reality: Context-awareness in augmented reality},
  author={Grubert, Jens and Langlotz, Tobias and Zollmann, Stefanie and Regenbrecht, Holger},
  journal={IEEE transactions on visualization and computer graphics},
  volume={23},
  number={6},
  pages={1706--1724},
  year={2016},
  publisher={IEEE}
}

@article{he2024adaptui,
  title={AdapTUI: Adaptation of Geometric-Feature-Based Tangible User Interfaces in Augmented Reality},
  author={He, Fengming and Hu, Xiyun and Qian, Xun and Zhu, Zhengzhe and Ramani, Karthik},
  journal={Proceedings of the ACM on Human-Computer Interaction},
  volume={8},
  number={ISS},
  pages={44--69},
  year={2024},
  publisher={ACM New York, NY, USA}
}

@inproceedings{lindlbauer2019context,
  title={Context-aware online adaptation of mixed reality interfaces},
  author={Lindlbauer, David and Feit, Anna Maria and Hilliges, Otmar},
  booktitle={Proceedings of the 32nd annual ACM symposium on user interface software and technology},
  pages={147--160},
  year={2019}
}

@inproceedings{cheng2021semanticadapt,
  title={Semanticadapt: Optimization-based adaptation of mixed reality layouts leveraging virtual-physical semantic connections},
  author={Cheng, Yifei and Yan, Yukang and Yi, Xin and Shi, Yuanchun and Lindlbauer, David},
  booktitle={The 34th Annual ACM Symposium on User Interface Software and Technology},
  pages={282--297},
  year={2021}
}

@inproceedings{wang2020capturar,
  title={CAPturAR: An augmented reality tool for authoring human-involved context-aware applications},
  author={Wang, Tianyi and Qian, Xun and He, Fengming and Hu, Xiyun and Huo, Ke and Cao, Yuanzhi and Ramani, Karthik},
  booktitle={Proceedings of the 33rd Annual ACM Symposium on User Interface Software and Technology},
  pages={328--341},
  year={2020}
}

@inproceedings{qian2022scalar,
  title={Scalar: Authoring semantically adaptive augmented reality experiences in virtual reality},
  author={Qian, Xun and He, Fengming and Hu, Xiyun and Wang, Tianyi and Ipsita, Ananya and Ramani, Karthik},
  booktitle={Proceedings of the 2022 CHI Conference on Human Factors in Computing Systems},
  pages={1--18},
  year={2022}
}

@inproceedings{nuernberger2016snaptoreality,
  title={Snaptoreality: Aligning augmented reality to the real world},
  author={Nuernberger, Benjamin and Ofek, Eyal and Benko, Hrvoje and Wilson, Andrew D},
  booktitle={Proceedings of the 2016 CHI conference on human factors in computing systems},
  pages={1233--1244},
  year={2016}
}

@inproceedings{chen2020augmenting,
	title        = {Augmenting Static Visualizations with PapARVis Designer},
	author       = {Zhu-Tian, Chen and Tong, Wai and Wang, Qianwen and Bach, Benjamin and Qu, Huamin},
	year         = 2020,
	booktitle    = {Proceedings of the 2020 CHI Conference on Human Factors in Computing Systems},
	pages        = {1--12}
}

@InProceedings{DBLP:conf/chi/HuangQWPSCRQ21,
  author    = {Gaoping Huang and Xun Qian and Tianyi Wang and Fagun Patel and Maitreya Sreeram and Yuanzhi Cao and Karthik Ramani and Alexander J. Quinn},
  booktitle = {Proc. of CHI},
  title     = {{AdapTutAR: An Adaptive Tutoring System for Machine Tasks in Augmented Reality}},
  year      = {2021},
  address   = {New York},
  pages     = {417:1--417:15},
  publisher = {{ACM}},
  bibsource = {dblp computer science bibliography, https://dblp.org},
  doi       = {10.1145/3411764.3445283},
  ranking   = {rank2},
  url       = {https://doi.org/10.1145/3411764.3445283},
}

@InProceedings{DBLP:conf/ismar/TaharaSNI20,
  author    = {Tomu Tahara and Takashi Seno and Gaku Narita and Tomoya Ishikawa},
  booktitle = {Proc. of ISMAR},
  title     = {{Retargetable {AR:} Context-aware Augmented Reality in Indoor Scenes based on 3D Scene Graph}},
  year      = {2020},
  address   = {Los Alamitos},
  pages     = {249--255},
  publisher = {{IEEE} Computer Society},
  bibsource = {dblp computer science bibliography, https://dblp.org},
  doi       = {10.1109/ISMAR-Adjunct51615.2020.00072},
  ranking   = {rank2},
  url       = {https://doi.org/10.1109/ISMAR-Adjunct51615.2020.00072},
}

@InProceedings{DBLP:conf/chi/FenderHA018,
  author    = {Andreas Fender and Philipp Herholz and Marc Alexa and J{\"{o}}rg M{\"{u}}ller},
  booktitle = {Proc. of CHI},
  title     = {{OptiSpace: Automated Placement of Interactive 3D Projection Mapping Content}},
  year      = {2018},
  address   = {New York},
  pages     = {269},
  publisher = {{ACM}},
  bibsource = {dblp computer science bibliography, https://dblp.org},
  doi       = {10.1145/3173574.3173843},
  ranking   = {rank2},
  url       = {https://doi.org/10.1145/3173574.3173843},
}

@InProceedings{DBLP:conf/uist/FenderLHA017,
  author    = {Andreas Fender and David Lindlbauer and Philipp Herholz and Marc Alexa and J{\"{o}}rg M{\"{u}}ller},
  booktitle = {Proc. of UIST},
  title     = {{HeatSpace: Automatic Placement of Displays by Empirical Analysis of User Behavior}},
  year      = {2017},
  address   = {New York},
  pages     = {611--621},
  publisher = {{ACM}},
  bibsource = {dblp computer science bibliography, https://dblp.org},
  doi       = {10.1145/3126594.3126621},
  ranking   = {rank2},
  url       = {https://doi.org/10.1145/3126594.3126621},
}

@article{DBLP:journals/tvcg/ChenS0WQW20,
  author       = {Chen Zhu-Tian and
                  Yijia Su and
                  Yifang Wang and
                  Qianwen Wang and
                  Huamin Qu and
                  Yingcai Wu},
  title        = {{MARVisT: Authoring Glyph-Based Visualization in Mobile Augmented Reality}},
  journal      = {{IEEE} Trans. Vis. Comput. Graph.},
  volume       = {26},
  number       = {8},
  pages        = {2645--2658},
  year         = {2020},
  url          = {https://doi.org/10.1109/TVCG.2019.2892415},
  doi          = {10.1109/TVCG.2019.2892415},
  bibsource    = {dblp computer science bibliography, https://dblp.org}
}

@inproceedings{evangelista2022auit,
  title={Auit--the adaptive user interfaces toolkit for designing xr applications},
  author={Evangelista Belo, Jo{\~a}o Marcelo and Lystb{\ae}k, Mathias N and Feit, Anna Maria and Pfeuffer, Ken and K{\'a}n, Peter and Oulasvirta, Antti and Gr{\o}nb{\ae}k, Kaj},
  booktitle={Proceedings of the 35th Annual ACM Symposium on User Interface Software and Technology},
  pages={1--16},
  year={2022}
}

@software{figma_design_2026,
  author       = {{Figma, Inc.}},
  title        = {Figma},
  year         = {2026},
  url          = {https://www.figma.com/},
  note         = {Collaborative interface design tool.}
}

@article{zhu2023rl,
  title={RL-L ABEL: A Deep Reinforcement Learning Approach Intended for AR Label Placement in Dynamic Scenarios},
  author={Zhu-Tian, Chen and Chiappalupi, Daniele and Lin, Tica and Yang, Yalong and Beyer, Johanna and Pfister, Hanspeter},
  journal={IEEE transactions on visualization and computer graphics},
  volume={30},
  number={1},
  pages={1347--1357},
  year={2023},
  publisher={IEEE}
}

@incollection{hutchins1986direct,
  title={Direct manipulation interfaces},
  author={Hutchins, Edwin L and Hollan, James D and Norman, Donald A},
  booktitle={User centered system design},
  pages={87--124},
  year={1986},
  publisher={CRC Press}
}

@book{foley1996computer,
  title={Computer graphics: principles and practice},
  author={Foley, James D},
  volume={12110},
  year={1996},
  publisher={Addison-Wesley Professional}
}

@inproceedings{whitlock2018interacting,
  title={Interacting with distant objects in augmented reality},
  author={Whitlock, Matt and Harnner, Ethan and Brubaker, Jed R and Kane, Shaun and Szafir, Danielle Albers},
  booktitle={2018 IEEE Conference on Virtual Reality and 3D User Interfaces (VR)},
  pages={41--48},
  year={2018},
  organization={IEEE}
}

@inproceedings{ren2026periphar,
  title={PeriphAR: Fast and Accurate Real-World Object Selection with Peripheral Augmented Reality Displays},
  author={Ren, Yutong and Reddy, Arnav and Nebeling, Michael},
  booktitle={Proceedings of the 2026 CHI Conference on Human Factors in Computing Systems},
  pages={1--16},
  year={2026}
}

@article{lee2018interaction,
  title={Interaction methods for smart glasses: A survey},
  author={Lee, Lik-Hang and Hui, Pan},
  journal={IEEE access},
  volume={6},
  pages={28712--28732},
  year={2018},
  publisher={IEEE}
}

\appendix
\newpage
\section*{\final{APPENDIX}}
\section{\final{Representative Ranking Sanity-Check Case}}
\label{sec:ranking-candidates}

\begin{figure*}[b]
    \centering
    \includegraphics[width=\textwidth]{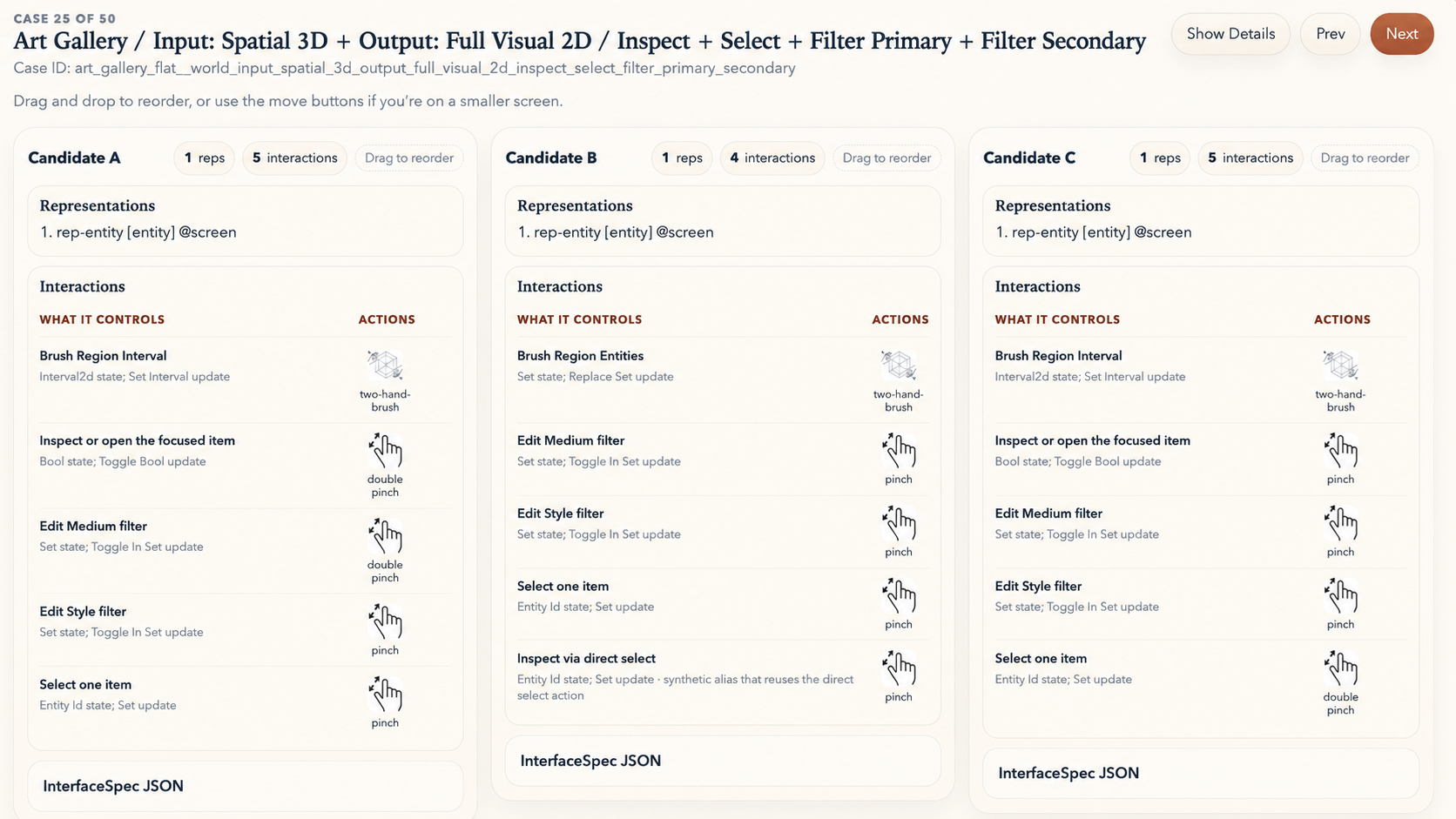}
    \caption{\final{Representative candidate set for Case 25 in the internal ranking sanity check. The art-gallery case uses spatial 3D input and full visual 2D screen output for inspection, single- and multi-selection, and primary and secondary filtering. The evaluation interface presents three anonymized candidate proxy UIs side by side.}}
    \Description{Evaluation interface for Case 25 comparing three anonymized candidate proxy user interfaces for a flat art-gallery scene with 3D-gesture input and screen output.}
    \label{fig:rank-candidate-ui}
\end{figure*}

\final{Figure~\ref{fig:rank-candidate-ui} presents one representative packet from the 50-case internal ranking sanity check. We selected Case~25, which uses a flat art-gallery scene with spatial 3D input and full visual 2D screen output. Its ordered intents cover inspection, selection, and primary and secondary filtering.}

\final{The evaluation interface preserves the view presented to the coders: three anonymized candidate proxy UIs (A--C) are shown side by side with their representation choices and gesture-to-intent mappings. 
The participant was asked to ranked these three options based on their personal preference.
This example illustrates how valid candidates for the same scene, device profile, and intent specification can differ in interaction assignment and state-update operations.}

\final{The complete scene, device capability, intent, and \texttt{InterfaceSpec} JSON for this case, together with the remaining ranking packets, is provided in \texttt{ranking\_\allowbreak sanity\_\allowbreak check\_\allowbreak supplement.zip} in the supplementary materials.}

\end{document}